\documentclass[11pt,a4paper]{article}
\usepackage{t1enc}
\usepackage[latin1]{inputenc}
\usepackage[english]{babel}
\normalfont
\usepackage{amsmath}
\usepackage{yfonts}
\usepackage{bbm}
\usepackage{bm}
\usepackage{wasysym}
\usepackage{amssymb}
\usepackage{mathrsfs}
\usepackage{pifont}
\usepackage{hyperref}
\usepackage{cite}

\newcommand{\be}[0]{\begin{equation}}
\newcommand{\ee}[0]{\end{equation}}

\numberwithin{equation}{section}

\begin{document}

\vspace*{-1cm}
\thispagestyle{empty}
\begin{flushright}
LPTENS 11/35
\end{flushright}
\vspace*{1.5cm}

\begin{center}
{\Large 
{\bf The anomaly line bundle of the self-dual field theory}}
\vspace{1.5cm}

{\large Samuel Monnier}%
\vspace*{0.5cm}

Laboratoire de Physique Th\'eorique de l'\'Ecole 
Normale Sup\'erieure\footnote{Unit\'e mixte de recherche (UMR 8549) du CNRS  et de l'ENS, associ\'ee \`a l'Universit\'e  Pierre et Marie Curie et aux f\'ed\'erations de recherche FR684 et FR2687.} 
\\
CNRS UMR 8549 \\
24 rue Lhomond, 75231 Paris Cedex 05, France \\ \vspace{.5cm}

Laboratoire de Physique Théorique et Hautes \'Energies \\
CNRS UMR 7589 and Université Pierre et Marie Curie - Paris 6 \\
4 place Jussieu, 75252 Paris cedex 05, France\\ \vspace{.5cm}
monnier@lpt.ens.fr 
\\ 
\vspace*{1cm}

{\bf Abstract}
\end{center}

In this work, we determine explicitly the anomaly line bundle of the abelian self-dual field theory over the space of metrics modulo diffeomorphisms, including its torsion part. Inspired by the work of Belov and Moore, we propose a non-covariant action principle for a pair of Euclidean self-dual fields on a generic oriented Riemannian manifold. The corresponding path integral allows one to study the global properties of the partition function over the space of metrics modulo diffeomorphisms. We show that the anomaly bundle for a pair of self-dual fields differs from the determinant bundle of the Dirac operator coupled to chiral spinors by a flat bundle that is not trivial if the underlying manifold has middle-degree cohomology, and whose holonomies are determined explicitly. We briefly sketch the relevance of this result for the computation of the global gravitational anomaly of the self-dual field theory, that will appear in another paper.

\newpage

\tableofcontents

\section{Introduction and summary}

The abelian self-dual field theory is a notoriously subtle quantum field theory. Its importance stems from its appearance in the field content of type IIB supergravity, on the worldvolume of the fivebranes in type IIA and heterotic $E_8 \times E_8$ string theories, on the world volume of the M5-brane, as well as in the effective description of the Coulomb branch of the conjectural (2,0) superconformal field theories in dimension 6. This paper realizes the first step of a program whose aim is to compute the global gravitational anomaly of the self-dual field theory. Given that the local and global gravitational anomalies of the chiral fermionic theories were computed in the mid 80's \cite{AlvarezGaume:1983ig, Witten:1985xe}, the fact that this problem is still open underlines the poor understanding we have of the self-dual field theory, especially when it is considered on non-trivial Riemannian manifolds. The latter setting is especially important for the computation of the low energy effective actions of string compactifications, to which Euclidean five-brane instantons contribute (see for instance \cite{Becker:1995kb, Witten:1996bn, Dijkgraaf:2002ac, Tsimpis:2007sx, Donagi:2010pd, Alexandrov:2010ca}). In this paper, we will consider the general theory of an abelian self-dual $2\ell$-form gauge theory in dimension $4\ell+2$. 
\\

Let us recall some basic facts about anomalies. Consider an Euclidean quantum field theory over a Riemannian manifold $M$. The metric on $M$ can be seen as an external parameter of the theory. In principle, we expect the partition function of the theory to be a function over the space $\mathcal{M}$ of metrics on $M$ modulo the group $\mathcal{D}$ of diffeomorphisms. For theories suffering from a gravitational anomaly, this is not quite true: the partition function is only the section of a complex line bundle $\mathscr{A}$ over $\mathcal{M}/\mathcal{D}$, that we will call the anomaly bundle. This bundle naturally comes with a Hermitian structure and a connection. The anomaly bundle can be topologically non-trivial; its class in the topological Picard group of $\mathcal{M}/\mathcal{D}$ is the \emph{topological anomaly}. Even if $\mathscr{A}$ is topologically trivial, it might not come with a natural trivialization. Indeed, the curvature and the holonomies of its connection might not vanish, and the theory displays respectively a \emph{local} or a \emph{global anomaly}. If these three types of anomalies vanish, then the partition function is the section of a bundle equipped with a flat connection with trivial holonomies, so can naturally be identified with an honest function over $\mathcal{M}/\mathcal{D}$. \footnote{Note that it is sufficient for the local and global anomalies to vanish for the anomaly bundle to be topologically trivial. This is why the topological anomaly is rarely mentioned explicitly in the literature. In our program, its interest stems from the handle it offers for the computation of the global anomaly.} The main reason for the study of gravitational anomalies comes from the fact that we expect low energy field theory limits of quantum gravity to be free of gravitational anomalies, so the study of the latter provides interesting information about what type of quantum field theories can arise in this limit.

The problem of computing gravitational anomalies was essentially solved by Alvarez-Gaumé and Witten in \cite{AlvarezGaume:1983ig} for local anomalies, and by Witten in \cite{Witten:1985xe} for global anomalies of chiral fermionic theories. These works were reformulated mathematically in terms of the index theory for families of Dirac operators by Bismut and Freed in \cite{MR853982, MR861886}, using the fact that the anomaly bundles of chiral fermionic theories happen to coincide with determinant bundles of Dirac operators. In \cite{MR853982, MR861886}, the authors were able to construct explicitly the Hermitian structures and connections on such determinant bundles, thereby recovering and generalizing the formulas of \cite{AlvarezGaume:1983ig, Witten:1985xe} for the local and global anomalies. \\

Concerning the self-dual field theory, it was noticed that a pair of self-dual fields is related to the Dirac operator coupled to chiral spinors, that we will call $D$. Using this insight, a formula for the local anomaly was proposed in \cite{AlvarezGaume:1983ig}, and was later checked in \cite{PhysRevLett.63.728} using an action principle for the self-dual field theory. This relation to $D$ can also be seen in the geometric quantization approach to the self-dual field theory \cite{Monnier:2010ww}. These works all show that the anomaly bundle of a pair of self-dual fields carries a connection whose curvature coincides with the curvature of the natural connection on the inverse of the determinant bundle $\mathscr{D}$ of $D$. We can deduce that modulo torsion, the anomaly bundle for a pair of self-dual fields coincides with $\mathscr{D}^{-1}$. 

The case of the global anomaly is much more subtle, because the latter is sensitive to torsion. A formula was proposed in \cite{Witten:1985xe}, in the case when the $4\ell+2$-dimensional manifold $M$ has no cohomology in degree $2\ell+1$, using again the relation to the Dirac operator coupled to chiral spinors. In hindsight \cite{MR861886}, this formula describes the holonomies of $\mathscr{D}^{-1}$, and it was pointed out in \cite{Witten:1985xe} that the presence of cohomology in degree $2\ell+1$ should invalidate it. 

We will see in this paper that this is indeed the case. When there is non-trivial cohomology in degree $2\ell+1$, our main result is that the anomaly bundle $\mathscr{A}^\eta$ for a pair of self-dual fields differs from $\mathscr{D}^{-1}$ by a flat bundle $\mathscr{F}^\eta$, whose holonomies can be computed explicitly. \footnote{$\eta$ is a discrete parameter on which the self-dual field theory depends and that can be identified with a theta characteristic.} Practically, we describe a certain family of (topological) line bundles on $\mathcal{M}/\mathcal{D}^{(2)}$. \footnote{Here $\mathcal{D}^{(2)}$ is a certain subgroup of the group of diffeomorphisms of $M$.} These bundles are pull-backs from bundles over a finite-dimensional modular variety $\mathcal{T}^{(2)}$, describing the polarizations of the intermediate Jacobian of $M$ modulo the action of $\mathcal{D}^{(2)}$. Fortunately, the bundles $\mathscr{D}^{-1}$ and $\mathscr{A}^\eta$ belong to this family, and the topological Picard group of $\mathcal{T}^{(2)}$ has been recently computed \cite{Sato01012010, 2009arXiv0908.0555P}. This allows us to get a complete description of the bundles in this family and to compute explicitly the holonomies of $\mathscr{F}^\eta$.

Topologically, $\mathscr{A}^\eta$ is the pull-back from $\mathcal{T}^{(2)}$ of the square of the theta bundle, of which the Siegel theta constant with characteristic $\eta$ is a section. Given the previous appearances of theta functions in the literature about the self-dual field \cite{Witten:1996hc, Henningson:1999dm, Dolan:1998qk, Dijkgraaf:2002ac, Belov:2006jd, Alexandrov:2010ca, Monnier:2010ww}, this comes as no surprise. However we would like to point out that this fact is far from being trivial. The theta bundle describes the anomaly due to the zero modes of the self-dual field. In principle, the non-zero modes can contribute as well to the topological anomaly, as it is the case in general for anomalous chiral fermionic theories. In the case of the self-dual field, the one-loop determinant has the very special property that it never vanishes over $\mathcal{M}/\mathcal{D}^{(2)}$, and hence is the section of a topologically trivial bundle. This is the crucial fact that allows us to study the anomaly bundle over the finite-dimensional space $\mathcal{T}^{(2)}$.

Given that the holonomies of $\mathscr{D}^{-1}$ are known from the work of Bismut-Freed \cite{MR861886}, our result about the holonomies of $\mathscr{F}^\eta$ in principle solves the problem of computing the global gravitational anomaly of a pair of self-dual fields. In practice, the formula we obtain here is difficult to use to check anomaly cancellation. Also, the problem of the computation of the global anomaly of a single self-dual field is left open: it is indeed a very non-trivial task to take the square root of a set of holonomies. These are issues that we hope to address in a future paper, in which the link with the work of Hopkins and Singer \cite{hopkins-2005-70} will also become apparent. \\

In order to identify the anomaly bundle $\mathscr{A}^\eta$, we need a global definition (over $\mathcal{M}/\mathcal{D}^{(2)}$) of the partition function, for an arbitrary choice of $4\ell+2$-dimensional oriented Riemannian manifold $M$. Finding such a definition turns out to be non-trivial. Geometric quantization methods \cite{Witten:1996hc, Belov:2006jd,Monnier:2010ww} apply to arbitrary Riemannian manifolds. They provide the Hermitian structure on $\mathscr{A}^\eta$ (equivalently the norm of the partition function as a function over $\mathcal{M}/\mathcal{D}^{(2)}$) and a local form of the natural connection on $\mathscr{A}^\eta$. This data does \emph{not} contain torsion information about $\mathscr{A}^\eta$.

Alternatively, a global way to construct the partition function is via path integration. Performing explicitly the path integration is not a hopeless task, because the abelian self-dual field is in principle a free theory. Quite a few actions have been proposed for the self-dual field (see for instance \cite{Henneaux:1988gg, McClain:1990sx, Pasti:1995tn, Pasti:1996vs, Devecchi:1996cp, Belov:2006jd}). The original action proposed by Henneaux-Teitelboim takes advantage of the fact that there exists a time-like Killing vector on Lorentzian manifolds, so it is not clear how it can be formulated on an arbitrary Riemannian manifold. The action of Pasti, Sorokin and Tonin \cite{Pasti:1995tn, Pasti:1996vs} is covariant, but it requires as well the existence of a non-vanishing vector field, whose existence is not guaranteed on a generic Riemannian manifold. Other covariant actions \cite{McClain:1990sx, Devecchi:1996cp} involve an infinite number of auxiliary fields, what makes the study of their path integral delicate \cite{Devecchi:1996cp}. 

The formalism we found most suitable is the one developed by Belov and Moore in \cite{Belov:2006jd}. Inspired by their work, we define an action for a pair of self-dual fields, and show that the associated path integral yields a partition function that is completely consistent with what we already know about the anomaly bundle. More precisely, we show how to recover the curvature of the anomaly bundle, or equivalently the local anomaly of the self-dual field theory. In the remainder of the paper, we assume that the partition function obtained from this action is the correct one for a pair of self-dual fields and extract from it global information about the anomaly bundle.

The Belov-Moore action was originally formulated for a single self-dual field. The action depends on a choice of Lagrangian subspace in the space of forms of degree $2\ell+1$ on $M$. Somewhat surprisingly, we have not been able to find the same consistent picture for the partition function obtained for a single self-dual field from the original Belov-Moore action. We believe that this problem can be traced back to a more fundamental question: what are the off shell degrees of freedom of a single self-dual field on an arbitrary Riemannian manifold? As we discuss in Section \ref{SecChoiceLag}, this question has no obvious answer, which might suggest that an Euclidean single self-dual field on a generic manifold cannot be formulated as a Gaussian Lagrangian theory.

In consequence, we only obtain global information for the anomaly bundle of \emph{a pair} of self-dual fields. Fortunately, the detailed knowledge of the family of bundles mentioned above allows us to single out an essentially unique square root for the anomaly bundle of a single self-dual field, see Section \ref{SecAnSingSDF} for a more precise statement. Note that a discussion about the anomaly bundle of the (2,0) superconformal theories in six dimensions appeared recently in \cite{Henningson:2010rc}.\\


The paper is organized as follows. We review the formalism of Belov-Moore in Section \ref{SecActBM} and define our action for a pair of self-dual fields in Section \ref{SecActPSDF}. In Section \ref{SecPathInt}, we perform the path integration. The latter decomposes into a sum over topological sectors (Section \ref{SecIntMassless}) and an integration over the non-zero modes (Section \ref{SecIntMassive}). We then show how to recover the local anomaly from our formula for the partition function (Section \ref{SecFPartSecLi}) and discuss informally the problems arising with actions for a single self-dual field on arbitrary Riemannian manifolds (Section \ref{SecChoiceLag}). Section \ref{SecModGeom} is a hopefully pedagogical introduction to modular geometry, factors of automorphy and their relations to line bundles. We also discuss recent and less well-known results providing a classification of topological line bundles over certain modular varieties. In Section \ref{SecAnom}, we characterize the anomaly bundles of a pair of self-dual fields (Section \ref{SecClassAn}) and of a single self-dual field (Section \ref{SecAnSingSDF}). We also discuss to which extent the information gained on bundles over the finite dimensional modular varieties apply to their pullback over $\mathcal{M}/\mathcal{D}^{(2)}$ (Section \ref{SecModSpaces}). In Section \ref{SecHolFormPSD} we briefly sketch how these results allow to extract a formula for the global gravitational anomaly of a pair of self-dual fields. Finally, in the appendix, we exhibit examples of 6-manifolds whose mapping class groups surjects on $Sp(2n,\mathbbm{Z})$, as they appear in an argument of Section \ref{SecModSpaces}.

\section{An action for a pair of self-dual fields in Euclidean signature}

\label{SecClassNCAct}

In this section, we propose an action for a pair of self-dual fields on a compact Riemannian manifold. This construction is strongly inspired by the work of Belov and Moore \cite{Belov:2006jd}. The action includes the coupling to a self-dual gauge field (the holographic Chern-Simons field of \cite{Witten:1996hc, Belov:2006jd}), as well as a ``characteristic'', a discrete parameter on which the quantum theory is known to depend. We will see in Section \ref{SecPathInt} that the path integral over this action correctly reproduces the known features of the partition function of a pair of self-dual fields. We also compute the value of the action on harmonic configurations of the self-dual field in Section \ref{SecRestHarm}.

\subsection{The action of Belov and Moore}

\label{SecActBM} 

Consider a compact oriented $4\ell + 2$-dimensional Riemannian manifold $M$, with $\ell$ an integer. The intersection product endows the space $\Omega^{2\ell + 1}(M,\mathbbm{R})$ of real valued $2\ell + 1$-forms on $M$ with a symplectic structure $\omega$:
\be
\omega(R,T) = 2\pi \int_M R \wedge T \;. 
\ee
The Hodge star operator $\ast$ coming from the metric on $M$ squares to $-\mathbbm{1}$ on $\Omega^{2\ell + 1}(M)$ and defines a complex structure. These two structures are compatible, turning $\Omega^{2\ell + 1}(M,\mathbbm{R})$ into a K\"ahler manifold. They pass to the space of harmonic $2\ell + 1$-forms on $M$, which we will identify with the cohomology group $H^{2\ell+1}(M,\mathbbm{R})$. The dimension of $H^{2\ell+1}(M,\mathbbm{R})$ will be denoted by $n$ in the following. To lighten the notation, we will drop the explicit dependence on $M$: $\Omega^{p} := \Omega^{p}(M,\mathbbm{R})$, and so forth. We have the Hodge decomposition $\Omega^{2\ell + 1} = H^{2\ell+1} \oplus \Omega_{\rm ex}^{2\ell + 1} \oplus \Omega_{\rm coex}^{2\ell + 1}$, where the last two summands denote the space of exact and co-exact forms, respectively. We will make a choice of Lagrangian decomposition of the lattice $\Lambda$ of harmonic $2\ell+1$-forms with integral periods: $\Lambda = \Lambda_1 \oplus \Lambda_2$. 
The sublattices $\Lambda_1$ and $\Lambda_2$ generate two Lagrangian subspaces $V^0_1$ and $V^0_2$ of $H^{2\ell+1}$.

Belov and Moore proposed an action for a single self-dual $2\ell$-form field in \cite{Belov:2006jd}, Section 7. They instruct to pick a Lagrangian subspace $V$ in $\Omega^{2\ell + 1}$ containing $V_2^0$. Its orthogonal complement with respect to $\omega(\,.\,,\ast\,.\,)$ is written $V^\perp$. The degrees of freedom for a single self-dual field are described by an element $R \in V_1^0 \oplus \Omega^{2\ell+1}_{\rm ex}$. \footnote{We are here effectively describing the self-dual field by its field strength, thereby ignoring the possible Wilson line degrees of freedom. This is justified by the fact that the latter have no effect on the computation of the partition function.} The Euclidean action then reads
\begin{equation}
\label{EqActBM}
S_V(R) = \pi \int_M \big( R^\perp \wedge \ast R^\perp + i R^\parallel \wedge R^\perp \big)\;,
\end{equation}
where $R^\parallel$ and $R^\perp$ are the components of $R$ along $V$ and $V^\perp$. They show that the equations of motion are those of a $2\ell$-form field $C$ with a $2\ell+1$-form self-dual field strength $F = R^\perp + i \ast R^\perp$. As $F$ is independent of  $R^\parallel$, this theory enjoys an extra gauge symmetry allowing to add an arbitrary exact form in $V$ to $R$. They also show that on a manifold of Lorentzian signature, one can recover the Henneaux-Teitelboim action \cite{Henneaux:1988gg} as a special case. 

Note that we actually have a whole family of actions, parameterized by the choice of a Lagrangian subspace $V$ in $\Omega^{2\ell+1}$. While all the members of this very large family yield the correct classical equations of motion, we have not been able to find a consistent interpretation for the quantum partition functions they yield upon path integration. This seems to be related to the fact that partition function of the Euclidean self-dual field theory is not (or at least not in an obvious way) of the Gaussian form. This very puzzling issue will be developed in Section \ref{SecChoiceLag}. 

Instead, we will see that the formalism of Belov and Moore can be adapted to construct an action for a pair of self-dual fields whose path integral yields a partition function consistent with what we already know of the self-dual field theory. The aim of the next section is to introduce this action.

\subsection{The action for a pair of self-dual fields}

\label{SecActPSDF} 

For a generic (oriented) Riemannian manifold $M$, essentially two obvious types of Lagrangian subspaces of $\Omega^{2\ell+1}$ are available. Modulo completion with a Lagrangian subspace of $H^{2\ell+1}$, these are given by $\Omega_{\rm ex}^{2\ell+1}$ and $\Omega_{\rm coex}^{2\ell+1}$. In a nutshell, our proposal for the action of a pair of self-dual fields is a sum of the Belov-Moore actions based on these two natural Lagrangian subspaces.

Consider first the Lagrangian subspace of $\Omega^{2\ell+1}$ given by 
\be
\hat{V} := V_2^0 \oplus \Omega^{2\ell+1}_{\rm ex} \;.
\ee
The orthogonal complement of $\hat{V}$ is 
\be
\hat{V}^\perp = V_2^{0\perp} \oplus \Omega^{2\ell+1}_{\rm coex} \;,
\ee
where $V_2^{0\perp} = \ast V_2^0$ denotes the orthogonal complement of $V_2^0$ in $H^{2\ell+1}$. From this data, we can construct a Belov-Moore action $S_{\hat{V}}(\hat{R})$, given by \eqref{EqActBM}. Recall that $\hat{R} \in V_1^0 \oplus \Omega^{2\ell+1}_{\rm ex}$. From the expression of $\hat{V}^\perp$, we see that $\hat{R}^\perp \in V_2^{0\perp}$, and the action is completely independent of the exact component of $\hat{R}$. This can be explained by the extra gauge symmetry enjoyed by the Belov-Moore action, that allows one to add an arbitrary exact form in $\hat{V}$ to $\hat{R}$. As $\Omega^{2\ell+1}_{\rm ex} \subset \hat{V}$, the exact component of $\hat{R}$ is pure gauge.

To describe the second half of the action, we need to pick a reference metric $g_0$ on $M$, kept fixed in all arguments, in addition to the actual metric $g$. The Lagrangian subspace of interest to us is $\check{V} := V_2^0 \oplus \Omega^{2\ell+1}_{\rm coex,0}$, where $\Omega^{2\ell+1}_{\rm coex,0}$ denotes the space of $2\ell+1$-forms that are co-exact with respect to the metric $g_0$. Note that just like $\hat{V}$, $\check{V}$ is independent of $g$, although it depends on our choice of a reference metric. We define $\check{V}^\perp := \ast \check{V}$ (where $\ast$ is the Hodge star operator associated with $g$). We can build as well a Belov-Moore action $S_{\check{V}}(\check{R})$ out of this data. 
This time, as there are no exact forms belonging to $\check{V}$, the action has no extra gauge symmetry and all the components of the field $\check{R}$ are physical.

It will be useful to us to add linear terms to the action, in order to recover the quantum partition function that was derived by geometric quantization methods \cite{Belov:2006jd, Monnier:2010ww}. Let $Z$ by a harmonic self-dual $2\ell+1$-form field satisfying $\ast Z = -i Z$ and $\eta^2$ a harmonic $2\ell+1$-form in $V_2^0$. We define
\be
S_{V,{\rm lin}}(R,Z, \eta^2) = 2\pi  i \int_M (2 Z \wedge R^\perp + \eta^2 \wedge R) \;.
\ee
These terms depend only on the harmonic component of $R$. Their interpretation (and the seemingly strange notation for $\eta^2$) will be clarified in Sections \ref{SecRestHarm} and \ref{SecIntMassless}. Note that similar terms appeared already implicitly in \cite{Belov:2006jd}, see for instance equation (6.20) there.

We are now ready to define our action principle. The degrees of freedom of the system are described by a pair $C = (\hat{C}, \check{C})$ of $2\ell$-forms gauge fields. Their field strengths can be described by a pair $R = (\hat{R},\check{R})$ of closed $2\ell+1$ forms whose harmonic components lie in $V_1^0$. \footnote{To be precise, $C$ is a pair of twisted differential characters \cite{springerlink:10.1007/BFb0075216, hopkins-2005-70, Freed:2006yc}, not of $2\ell$-forms. This is what allows the field strengths to have non-trivial harmonic components. These differential characters are ``twisted'' because the harmonic components of their field strengths are not necessarily integral. The space of twisted differential characters whose field strengths are in a given cohomology class is an affine space modeled on $\Omega^{2\ell}$, so one can picture them as $2\ell$-forms after a reference differential character has been singled out.} The action we propose for a pair of self-dual fields reads 
\begin{equation}
\label{EqActBMGen}
S(R,Z,\eta^2) = S_{\hat{V}}(\hat{R}) + S_{\hat{V},{\rm lin}}(\hat{R},Z, \eta^2) + S_{\check{V}}(\check{R}) + S_{\check{V},{\rm lin}}(\check{R},Z, \eta^2) \;.
\end{equation}
We define a variational problem for the action \eqref{EqActBMGen} by allowing variations of $C$ by arbitrary pairs of $2\ell$-forms. Note also that we can generalize this action further by adding couplings to sources (see equation (7.14) in \cite{Belov:2006jd}), but this will not be needed for our purpose.

As $R$ varies only by exact forms, the linear terms $S_{\hat{V},{\rm lin}}(\hat{R},Z, \eta^2) + S_{\check{V},{\rm lin}}(\check{R},Z, \eta^2)$ do not contribute to the equations of motion. By Theorem 7.4 of \cite{Belov:2006jd}, \eqref{EqActBMGen} yields the correct equations of motion for a pair of self-dual fields with field strengths
\be
F = (\hat{R}^\perp + i \ast \hat{R}^\perp, \check{R}^\perp + i \ast \check{R}^\perp) \;.
\ee
The equations of motion force $F$ to be a pair of harmonic self-dual forms.

Let us insist that the action \eqref{EqActBMGen} should be construed as describing a single free Gaussian theory, even if it is defined in terms of two Belov-Moore actions for the components $\hat{R}$ and $\check{R}$. Indeed, we have not been able to find a consistent interpretation for the quantum partition functions obtained from these two Belov-Moore actions in isolation. In contrast, we will see in Section \ref{SecFPartSecLi} that the partition function obtained from the full action matches perfectly our expectations for the partition function of a pair of self-dual fields. This point will be discussed in more detail in Section \ref{SecChoiceLag}.

\subsection{Restriction of the action on harmonic forms}

\label{SecRestHarm}

Now we will compute the value of the action \eqref{EqActBMGen} for $(\hat{R},\check{R})$ a pair of harmonic forms, in an explicit coordinate system on $H^{2\ell+1}$. This result will be useful for the computation of the path integral, and should clarify the nature of the linear terms in the action \eqref{EqActBMGen}. Let us start by choosing a Darboux basis $\{\alpha_i, \beta^i\}_{i=1}^n$ of $\Lambda$, such that the set $\{\alpha_i\}$ generates $\Lambda_1$ and the set $\{\beta_i\}$ generates $\Lambda_2$. We have by definition
\be
\omega(\alpha_i, \alpha_j) = \omega(\beta^i,\beta^j) = 0 \;, \qquad \omega(\alpha_i,\beta^j) = 2 \pi \delta_i^j \;.
\ee
We introduce coordinates $a^i$, $b_i$ such that 
\be
\alpha_i = \frac{\partial}{\partial a^i} \;, \qquad \beta^i = \frac{\partial}{\partial b_i} \;.
\ee
The action of the Hodge star operator on $H^{2\ell+1}(M,\mathbbm{C})$ can be captured by a period matrix $(\tau_{ij})_{i,j=1}^n$. Indeed, the holomorphic coordinates with respect to this complex structure can be written uniquely as 
\be
\label{EqDefz}
z_i = \tau_{ij}a^i + b_i \;,
\ee 
where the matrix $\tau$ is symmetric and has a positive definite imaginary part. 
Let us define the metric $h_{ij} = -i(\tau - \bar{\tau})_{ij}$, as well as the inverse metric $h^{ij} = i\big((\tau-\bar{\tau})^{-1}\big)^{ij}$. When indices are omitted, the latter matrix is denoted by $h^{-1}$. 
Then 
\be
a^i = -ih^{ik}(z_k - \bar{z}_k) \;, \qquad b_i = i\bar{\tau}_{ij}h^{jk}z_k - i\tau_{ij}h^{jk}\bar{z}_k \;.
\ee
$\omega$ can is expressed as follows:
\be
\omega = 2\pi da^i \wedge db_i = -2\pi i h^{ij} dz_i \wedge d\bar{z}_j \;.
\ee

We now compute the matrix of $\ast$ in terms of $\tau$ in the Darboux basis. We write
\be
\ast \alpha_i = (\ast_{11})_{i}^{\;j} \alpha_j + (\ast_{21})_{ij} \beta^j \;, \qquad \ast \beta^i = (\ast_{12})^{ij} \alpha_j + (\ast_{22})^{i}_{\;j} \beta^j \;.
\ee
We also decompose $\tau = x + iy$, with $x$ and $y$ real matrices. From \eqref{EqDefz}, we deduce the expression for the holomorphic vectors:
\be
\label{EqDefZeta}
\zeta^i = -ih^{ij}(\alpha_j - \bar{\tau}_{jk} \beta^k) \;.
\ee
From $\ast \zeta_j = -i\zeta_j$, we get
\begin{align}
(\ast_{11}) =&\; -(xy^{-1}) = -(\tau + \bar{\tau})h^{-1} \;, \notag\\
(\ast_{12}) =&\; -(y^{-1}) = -2h^{-1} \;, \\
(\ast_{21}) =&\; y + (xy^{-1}x) = \frac{1}{2} \big(h + (\tau + \bar{\tau})h^{-1}(\tau + \bar{\tau}) \big) \;, \notag\\
(\ast_{22}) =&\; (y^{-1}x) = h^{-1}(\tau + \bar{\tau}) \;. \notag
\end{align}
Using this explicit form for $\ast$, we see that $V_2^{0\perp}$ is generated by $\{\gamma_j\}_{j=1}^n$, with
\be
\gamma_j =  \alpha_j - \frac{1}{2}(\tau + \bar{\tau})_{jk} \beta^k \;.
\ee
We can compute the components of $\alpha_k$ on $V^0_2$ and $V_2^{0\perp}$, respectively:
\begin{align}
(\alpha_k)_2 =&\; \frac{1}{2}(\tau + \bar{\tau})_{kj}\beta^j \;, \\
(\alpha_k)^\perp_2 =&\; \alpha_k - \frac{1}{2}(\tau + \bar{\tau})_{kj}\beta^j \;. \notag
\end{align}
We will also need
\be
\ast(\alpha_k)^\perp_2 = \frac{1}{2} h_{kj}\beta^j \;.
\ee
Now suppose $R$ is a harmonic form of degree $2\ell+1$. Write $R = r^i \alpha_i$, $Z = z_i \zeta^i$ and $\eta^2 = \eta^2_i \beta^i$. We get
\begin{align}
\label{EqFormExplTermAction}
\pi \int_M R^\perp \wedge \ast R^\perp = &\; \frac{\pi}{2} r^ih_{ij}r^j =  -\frac{\pi i}{2} r^i(\tau-\bar{\tau})_{ij}r^j \;, \notag \\
\pi i \int_M R^\parallel \wedge R^\perp = &\; -\frac{\pi i}{2} r^i (\tau + \bar{\tau})_{ij} r^j \;,\\
4\pi i \int_M Z \wedge R^\perp = &\; -2\pi i z_j r^j      \;, \notag \\
2\pi i \int_M \eta^2 \wedge R = &\; -2\pi i \eta^2_i r^i   \;. \notag
\end{align}
Writing now $\hat{R} = \hat{r}^i \alpha_i$ and $\check{R} = \check{r}^i \alpha_i$, we find for \eqref{EqActBMGen}:
\be
\label{EqActOnHarm}
S(R,Z,\eta^2) = - \big( \pi i \hat{r}^i \tau_{ij} \hat{r}^j + 2\pi i (z_i + \eta^2_i)\hat{r}^i \big) - \big( \pi i \check{r}^i \tau_{ij} \check{r}^j + 2\pi i (z_i + \eta^2_i)\check{r}^i \big) \;.
\ee
Note that $e^{-S}$ is the product of two copies of the well-known kernel of the classical Siegel theta function with argument $Z$. In this interpretation, $\eta^2$ is the component on $V_2^0$ of the characteristic of the theta function.

\section{Path integration}

\label{SecPathInt}

We now compute the quantum partition function by means of a path integration with the classical action \eqref{EqActBMGen}. The path integral can be split into a factor $N$ coming from the integration of the non-zero modes and a factor $\mathcal{Z}_{0}(Z,\eta^2)$ coming from the integration on the zero modes:
\be
\mathcal{Z}(Z,\eta^2) = N \mathcal{Z}_{0}(Z,\eta^2) \;.
\ee 
As the linear terms in the action \eqref{EqActBMGen} decouple from the non-zero modes, $N$ depends neither on $Z$ nor on $\eta^2$.

\subsection{Sum over topological sectors}

\label{SecIntMassless}

The zero modes of $R$ are pairs of harmonic forms in $V_1^0$. They label topologically distinct field strengths. In order to define the path integration on these massless modes, we have to prescribe which topological sectors (instantons) have to be incorporated in the quantum theory. In general, we expect the field strength of a gauge field to have integral periods, because it can be interpreted as the curvature of a connection (the gauge field) on the gauge bundle. So we would expect $\hat{R}$ and $\check{R}$ to project on $\Lambda_1$. As we will see in Section \ref{SecModGeom}, covariance with respect to the action of the mapping class group of $M$ forces us to consider a slightly more general case, in which we allow $\hat{R}$ and $\check{R}$ to project on a shifted lattice: $\hat{R}|_{H^{2\ell+1}}, \check{R}|_{H^{2\ell+1}} \in \Lambda_1 + \eta_1$, where $\eta_1 \in V_1^0$ is a constant vector. \footnote{This generalization is directly related to the numerous appearances of half-integral quantized gauge fields in string theory. We will see at the end of Section \ref{SecAnSingSDF} that consistency requires $\eta_1$ to be half-integral.} 

The integration over the zero modes reduces to an infinite double sum over $\Lambda_1 + \eta_1$ and using \eqref{EqActOnHarm} we get:
\begin{align}
\label{EqFPartZeroModes}
\mathcal{Z}_{0}(Z,\eta) = &\;  \sum_{\hat{R} \in \Lambda_1 + \eta_1} \exp \big(\pi i \hat{r}^i \tau_{ij} \hat{r}^j + 2\pi i (z_i + \eta^2_i)\hat{r}^i \big) \cdot \sum_{\check{R} \in \Lambda_1 + \eta_1} \exp \big(\pi i \check{r}^i \tau_{ij} \check{r}^j + 2\pi i (z_i + \eta^2_i)\check{r}^i \big) \notag \\
= &\; \big(\theta^\eta(Z,\tau)\big)^2 \;,
\end{align}
where now $\mathcal{Z}_0$ depends on the full vector $\eta = (\eta_1, \eta^2) \in H^{2\ell+1}$. $\mathcal{Z}_0$ is nothing but the square of a classical Siegel theta function. It is well-known that the instanton sum over the zero modes of the self-dual field produces a theta function \cite{Witten:1996hc, Henningson:1999dm, Dolan:1998qk, Dijkgraaf:2002ac}. The linear terms we introduced in the action find here their justification: they correctly reproduce the dependence of the partition function on the external parameters $Z$ and $\eta$.

\subsection{Path integral over the non-zero modes}

\label{SecIntMassive}

We compute now the path integral $N$ of the action on non-zero modes. As the action is Gaussian, the path integral can in principle be performed exactly. We are not concerned with the harmonic part of $R$ so we can take $C = (\check{C},\hat{C})$ to be a pair of ordinary $2\ell$-forms, and $(\hat{R}, \check{R}) = (d\check{C},d\hat{C})$. 

Before performing the path integration, it is useful to reformulate a bit the action. First, we saw that the action is independent of the exact component of $\hat{R}$. Therefore in the absence of harmonic component, the action for $\hat{R}$ vanishes identically.  Writing $(.,.)_0$ for $\omega(.,\ast_0 .)$, we have
\be
S(R) = \pi \int_M  \check{R}^\perp \wedge \left( \ast \check{R}^\perp - i \check{R}^\parallel \right) = \frac{1}{2} \omega(\check{R}, \ast \check{R}^\perp - i \check{R}^\parallel) = - \frac{1}{2}(\check{R}, \ast_0 \ast \check{R}^\perp - i \ast_0 \check{R}^\parallel)_0 \;.
\ee
In the second equality, we used the fact that $\check{R}^\parallel$ belongs to a Lagrangian subspace, and the fact that $\check{R}^\parallel$ and $\check{R}^\perp$ are orthogonal with respect to $\omega(.,\ast .)$. Now recall that $\check{R}^\parallel \in \Omega^{2\ell+1}_{\rm coex,0}$ and that $\check{R}^\perp \in \ast\Omega^{2\ell+1}_{\rm coex,0}$. Write $P^\parallel$ and $P^\perp$ the projectors on these subspaces. We see that the operator
\be
\tau_+ := -i \ast_0 \ast P^\perp - \ast_0 P^\parallel \;,
\ee
maps $\Omega^{2\ell+1}_{\rm ex}(M,\mathbbm{C})$ to itself and that
\be
S(R) = -\frac{i}{2}(\check{R},\tau_+ \check{R})_0 \;.
\ee
Note that when $g_0 = g$, then $\tau_+ = i\mathbbm{1}$. $\tau_+$ can be pictured as an infinite-dimensional analog of the period matrix $\tau$ describing the polarization on the space of harmonic forms. $\tau_+$ describe the polarization generated by $\ast$ on the infinite dimensional space $\Omega^{2\ell+1}_{\rm ex} \oplus \Omega^{2\ell+1}_{\rm coex,0}$, with respect to the reference polarization defined by $\ast_0$. It is an element of an infinite dimensional analogue of the Siegel upper-half plane, that was defined in Section 3.1 of \cite{Monnier:2010ww}. \footnote{We recently realized that similar ideas appeared a long time ago in \cite{Henningson:1999dm}.} 

$\check{R}$ is exact and we use the parameterization $\check{R} = d\check{C}$. If $\ast_0$ is the Hodge star operator associated to the reference metric $g_0$, then let us write $d^\dagger_0 := -\ast_0 d \ast_0$. By a slight abuse of notation, we write $\tau_+$ as well for the operator $d_0^\dagger \tau_+  (d_0^\dagger)^{-1}$ acting on $\Omega^{2\ell}_{\rm coex,0}$.
The final form of the action is then
\be
S(R) = -\frac{i}{2}(\check{C}, \tau_+ d_0^\dagger d \check{C})_0 \;.
\ee

The path integral reads formally
\be
\label{EqNormFactTemp1}
N = \frac{1}{{\rm Vol}\mathcal{\check{G}}{\rm Vol}\mathcal{\hat{G}}} \int_{\Omega^{2\ell}} D\check{C} \int_{\Omega^{2\ell}} D\hat{C} \exp \left( \frac{i}{2}(\check{C}, \tau_+ d_0^\dagger d \check{C})_0 \right) \;,
\ee
where ${\rm Vol}\mathcal{\check{G}}$ and ${\rm Vol}\mathcal{\hat{G}}$ are the ``volumes of the gauge groups''. The expression \eqref{EqNormFactTemp1} is meaningless until we define the measures $D\check{C}$ and $D\hat{C}$. Note however that the integrand is independent of $\hat{C}$, therefore whatever choice we make for $D\hat{C}$, the integral will cancel the factor ${\rm Vol}\mathcal{\hat{G}}$, which is computed with the same measure. We have therefore
\be
\label{EqNormFactTemp2}
N = \frac{1}{{\rm Vol}\mathcal{\check{G}}} \int_{\Omega^{2\ell}} D\check{C} \exp \left( \frac{i}{2}(\check{C}, \tau_+ d_0^\dagger d \check{C})_0 \right) \;.
\ee
The action functional is simple enough to make possible the following heuristic reasoning inspired by \cite{Pestun:2005rp}. It should be seen as a way of figuring out what is the correct definition of the measure of the path integration. No honest derivation of the path integral is possible for lack of a definition of this measure from first principles. Performing the infinite Gaussian integral in \eqref{EqNormFactTemp2}, we get the inverse of the square root of the determinant of $\tau_+ d_0^\dagger d$ restricted to the complement of its kernel, times the volume of its kernel, which consists of all the closed $2\ell$-forms:  
\be
N = \frac{1}{{\rm Vol}\mathcal{\check{G}}} \, {\rm det}\big( -i\tau_+ \big)^{-1/2} \, {\rm det}\big( d_0^\dagger d|_{\Omega^{2\ell}_{\rm coex}}\big)^{-1/2} \, {\rm Vol}\Omega^{2\ell}_{\rm cl} \;,
\ee
where $\det$ denotes the zeta-regularized determinant. We can write 
\be
\label{EqVolClFrm}
{\rm Vol}\Omega^{p}_{\rm cl} = {\rm Vol}H^p {\rm Vol}\Omega^{p}_{\rm ex} \;.
\ee
The bijection
\be
d: \Omega^{p}/\Omega^{p}_{\rm cl} \rightarrow \Omega^{p+1}_{\rm ex}
\ee 
implies the equality
\be
\label{EqVolExFrm}
{\rm Vol}\Omega^{p+1}_{\rm ex} = {\rm Vol}\Omega^{p}({\rm Vol}\Omega^{p}_{\rm cl})^{-1} {\rm det} \big( d_0^\dagger d|_{\Omega^{p}_{\rm coex}}\big)^{1/2} \;.
\ee
Combining equations \eqref{EqVolClFrm} and \eqref{EqVolExFrm} and performing the recursion, we get
\be
N = \frac{1}{{\rm Vol}\mathcal{\check{G}}} \, {\rm det}\big( -i \tau_+ \big)^{-1/2} \, \prod_{p = 0}^{2\ell} ({\rm Vol}\Omega^{p})^{(-1)^{p+1}} ({\rm Vol}H^p)^{(-1)^p} \left ( {\rm det}(d_0^\dagger d|_{\Omega^p_{\rm coex}}) \right )^{(-1)^{p+1}/2} \;.
\ee
The factors $({\rm Vol}\Omega^{p})^{(-1)^{p+1}}$ can be canceled by a proper definition of ${\rm Vol}\mathcal{\check{G}}$. The factors $({\rm Vol}H^p)^{(-1)^p}$ cannot, because they depend on $g_0$. However they can be properly regularized as the volume of the unit box defined by an integral basis of harmonic forms with respect to the metric $g_0$ \cite{Schwarz:1979ae}. We call the regularized volume ${\rm Vol}_0H^p$. We find therefore the regularized expression 
\be
\label{EqFinExprN}
N = {\rm det}\big( -i \tau_+ \big)^{-1/2} \, \prod_{p = 0}^{2\ell} ({\rm Vol}_0H^p)^{(-1)^p} \left ( {\rm det}(d_0^\dagger d|_{\Omega^p_{\rm coex}}) \right )^{(-1)^{p+1}/2} \;.
\ee

$N$ has the remarkable property that it never vanishes on the space of Riemannian metric on $M$. To see this, note that the eigenspace of zero modes of the Laplacian is the space of harmonic forms and that its dimension is fixed by cohomology. It is therefore not possible for a co-exact form to have zero eigenvalue. Moreover, the imaginary part of $\tau_+$ is positive definite, therefore $|N|$ never vanishes. This fact will have important consequences for the study of the anomaly bundle. \\

It is useful to write:
\be
\label{EqMesSR}
u(g) = \prod_{p = 0}^{2\ell} \left ( \left ( {\rm Vol}(H^p)^{-2} {\rm det}'(d^\dagger d|_{\Omega^p_{\rm coex}}) \right )^{(-1)^p} \right )^{1/2}\;.
\ee
Modulo the volume factors, $u(g)$ coincides with the Cheeger half-torsion of $M$ \cite{Branson2005, Monnier:2010ww}. Combining \eqref{EqFPartZeroModes}, \eqref{EqFinExprN} and \eqref{EqMesSR}, the full partition function reads
\be
\label{EqFullPartF}
\mathcal{Z}(Z,\eta)=  u^{-1}(g_0) \det(-i \tau_+)^{-1/2} (\theta^\eta(Z,\tau))^2\;.
\ee
In the following we will sometimes specialize to $Z = 0$ in order to avoid unnecessary complications in the discussion. The dependence of the partition function on $Z$ and the anomaly bundle over the intermediate Jacobian of which it is a section are already well-understood \cite{Witten:1996hc}. A technical point is that the theta functions with odd characteristic vanish at $Z = 0$, so we will restrict ourself to even characteristic. We write $\mathcal{Z}(\eta) := \mathcal{Z}(0,\eta)$ and $\theta^\eta(\tau) := \theta^\eta(0,\tau)$.

\subsection{The partition function as a section of a line bundle}

\label{SecFPartSecLi}

At first sight, the partition function \eqref{EqFullPartF} has several puzzling features. First, it clearly depends on the choice of the reference metric $g_0$. $g_0$ was used to pick a Lagrangian subspace in order to define the Belov-Moore action, but it is certainly not physical. Second, the norm of the partition function of the self-dual field theory was computed in \cite{Belov:2006jd} and in \cite{Monnier:2010ww} using two different methods and it was found that for a pair of self-dual fields,
\be
\label{EqNormFuncPart}
|\mathcal{Z}(Z,\eta)|= (\det h)^{1/2} u^{-1}(g) |\theta^\eta(Z,\tau)|^2 \;.
\ee
A naive computation of the norm of \eqref{EqFullPartF} by multiplying it with its complex conjugate contradicts this result.

This situation can be explained if we remember that the partition function of the self-dual field is not a function over the space $\mathcal{M}$ of Riemannian metrics on $M$, but rather the section of a certain line bundle. We will write $\mathscr{A}^\eta$ for the anomaly bundle of a pair of self-dual fields. This bundle is naturally endowed with a Hermitian structure and a compatible connection, that have been computed in \cite{Monnier:2010ww}, at least locally. The curvature of the connection gives the local gravitational anomaly of the self-dual field theory. We will show here how to recover the local anomaly from the knowledge of the partition function \eqref{EqFullPartF} and how to solve the puzzles above. 

To this end, it is useful to consider an infinite dimensional space $\tilde{\mathcal{C}}$ of polarizations of $\Omega^{2\ell+1}(M)$, in which the space of metrics on $M$ is mapped. This space was defined in Section 3.1 of \cite{Monnier:2010ww} (see also \cite{Henningson:1999dm}). For our purpose, $\tilde{\mathcal{C}}$ can be thought of as parameterizing the possible actions of Hodge star operators on $\Omega^{2\ell+1}(M)$, although only a subset of those is realized by Hodge star operators associated to actual metrics on $M$. $\tilde{\mathcal{C}}$ is an infinite dimensional analogue of the Siegel upper half-space; in particular, it is a contractible space on which $\{\tau,\tau_+\}$ are coordinates. The interest of considering $\tilde{\mathcal{C}}$ lies in the fact that it is a complex space, and that $\mathscr{A}^\eta$ is the pull-back to $\mathcal{M}$ of a holomorphic line bundle on $\tilde{\mathcal{C}}$ (that we will still call $\mathscr{A}^\eta$). In contrast, there is no obvious complex structure on $\mathcal{M}$. As a result, by studying $\mathscr{A}^\eta$ as a bundle over $\tilde{\mathcal{C}}$, we can use the powerful tools holomorphicity provides.  

Given a holomorphic line bundle $\mathscr{L}$ over a base $B$, a Hermitian structure on $\mathscr{L}$ in a given holomorphic trivialization can be written as
\be
\label{EqHermStrucKalPot}
(s_1,s_2) = \exp\left(-K\right) f_1 \bar{f}_2 \;,
\ee
where $f_1$ and $f_2$ are holomorphic functions representing the sections $s_1$ and $s_2$ in the chosen trivialization. $K$ is a real-valued function that encodes the Hermitian structure on $\mathscr{L}$. By analogy with Kähler geometry, we will call it the Kähler potential. There is a unique compatible connection that reads (in the chosen trivialization)
\be
\nabla_{\mathscr{L}} = d_B - \partial_B K \;,
\ee
where $d_B$ and $\partial_B$ are the differential and the Dolbeault operator on $B$. \footnote{The compatibility condition between the Hermitian structure and the connection reads $d_B(s_1,s_2) = (\nabla_{\mathscr{L}}s_1,s_2) + (s_1, \nabla_{\mathscr{L}}s_2)$.} The curvature of $\nabla_{\mathscr{L}}$ is then given by 
\be
\label{EqCurvFromKalPot}
R_{\mathscr{L}} = \partial_B \bar{\partial}_B K \;.
\ee
Given our knowledge of the partition function, our aim is to compute the local anomaly, i.e. the curvature $R_{\mathscr{\mathscr{A}^\eta}}$. To this end, it is useful to note that the anomaly bundle can be decomposed as $\mathscr{A}^\eta = \mathscr{A}^\eta_0 \otimes \mathscr{A}_+$, where the two components of the tensor product correspond to the zero modes and the non-zero modes. We will consider separately the contributions of $\mathscr{A}^\eta_0$ and of $\mathscr{A}_+$ to the Kähler potential of $\mathscr{A}^\eta$, and then extract the curvature from it.

\paragraph*{Zero modes} The zero modes contribute a factor $(\theta^\eta(\tau))^2$ to the partition function. We will see at the end of Section \ref{SecFacAut} that such ``theta constants'' are sections of a holomorphic line bundle $\mathscr{A}^\eta_0$ over $\tilde{\mathcal{C}}$ whose Kähler potential is given by $-\frac{1}{2}\ln \det h$. The contributions of $\mathscr{A}^\eta_0$ to the Kähler potential of $\mathscr{A}^\eta$ is therefore $-\ln \det h$.

\paragraph*{Non-zero modes} The contribution of the non-zero modes is more tricky to interpret. Let us go back to our Hermitian line bundle $\mathscr{L}$ and suppose that it is topologically trivial. Let $s$ be a holomorphic non-vanishing section of $\mathscr{L}$ and $x_0 \in B$ be a base point. Locally around $x_0$, there exists a unique holomorphic trivialization of $\mathscr{L}$ in which $s$ has the form 
$$
s_{x_0}(x) = |s(x_0)| f_{x_0}(x) \;,
$$
where $f_{x_0}(x)$ is a holomorphic function normalized so that $f_{x_0}(x_0) = 1$. From \eqref{EqHermStrucKalPot}, we see that in this trivialization, the Kähler potential vanishes at $x_0$. In terms of the norm of $s$, it reads explicitly
\be
\label{EqKahlPot}
K(x) = -\ln |s(x)|^2 + \ln |s(x_0)|^2 + \ln |f(x)|^2 \;.
\ee

The contribution of the non-zero modes to the path integral fits exactly into this picture. Recall that it reads
\be
N = u^{-1}(g_0) \det(-i \tau_+)^{-1/2} \;.
\ee
$u(g)$ is a strictly positive function and $\det(-i \tau_+)^{-1/2}$ is a holomorphic function on $\tilde{\mathcal{C}}$ which is equal to $1$ when $\tau_+ = i\mathbbm{1}$, that is when $g = g_0$. We therefore interpret the choice of a reference metric $g_0$ as a choice of trivialization of $\mathscr{A}_+$. The Kähler potential in the trivialization associated with $g_0$ reads
\be
K_+(g) = 2 \ln u(g) - 2 \ln u(g_0) - \ln |\det(-i \tau_+)|
\ee
according to \eqref{EqKahlPot}.

\paragraph{}The full Kähler potential reads
\be
\label{EqKalPotA}
K(g) = -\ln \det h\, + 2 \ln u(g) - 2 \ln u(g_0) - \ln |\det(-i \tau_+)| \;.
\ee
The norm of the partition function for a pair of self-dual fields is therefore
\be
\label{EqNormPartFunc}
|\mathcal{Z}(Z=0,\eta)|= \sqrt{\det h} \, u^{-1}(g)|\theta^\eta(\tau)|^2 \;,
\ee
in accordance with \eqref{EqNormFuncPart}. The connection on $\mathscr{A}^\eta$ is
\be
\label{EqConnAnomBundle}
\nabla_{\mathscr{A}^\eta} = d_{\tilde{\mathcal{C}}} + h^{ij} \partial_{\tilde{\mathcal{C}}} h_{ij} - 2\partial_{\tilde{\mathcal{C}}} \ln u + \frac{1}{2} \partial_{\tilde{\mathcal{C}}} \det(-i \tau_+) \;.
\ee
Its curvature reads
\be
\label{EqCurvAnomBundle}
R_{\mathscr{A}^\eta} = - \partial_{\tilde{\mathcal{C}}} \bar{\partial}_{\tilde{\mathcal{C}}} \ln \det h + 2 \partial_{\tilde{\mathcal{C}}} \bar{\partial}_{\tilde{\mathcal{C}}} \ln u \;.
\ee
This agrees with twice the expression derived in \cite{Monnier:2010ww} for the curvature of the anomaly bundle of a single self-dual field, using geometric quantization techniques. Therefore the quantum partition function we derived from our action displays the correct local gravitational anomaly.

Note that a computation looking superficially very similar to our computation of the partition function has been performed in \cite{Henningson:1999dm}. It should in principle be possible as well to recover the local anomaly from the expressions of \cite{Henningson:1999dm}. See also \cite{Dijkgraaf:2002ac} for yet another similar computation of the partition function.

\subsection{Is the free self-dual field a Gaussian theory?}

\label{SecChoiceLag}

We would like here to comment on some issues about actions for a single self-dual field on generic Riemannian manifolds. In the previous sections, we have presented an action principle for a pair of self-dual fields. The latter passed three important tests. 
\begin{enumerate}
	\item It correctly reproduces the classical equations of motion for self-dual fields.
	\item After the gauge symmetry has been taken into account, it displays the off shell degrees of freedom expected for a pair of self-dual fields, equivalent to the off shell degrees of freedom of a single ordinary abelian gauge field. While the off shell degrees of freedom are irrelevant to the classical theory, they are of course of prime importance for the quantum theorym, where for instance they determine the one-loop determinant of the partition function.
	\item We checked that the quantum partition function on any oriented Riemannian manifold is a section of a holomorphic line bundle, whose curvature is consistent with the known local anomaly.
\end{enumerate}

We have \emph{not} been able to find an action for a single self-dual field within the Belov-Moore formalism that satisfies these consistency conditions on an arbitrary oriented Riemannian manifold. It might be that we did not try hard enough, but we feel there is a deeper and more interesting explanation for this. Let us go back to our computation of the path integral. Our pair of self-dual fields was divided into two components $\hat{R}$ and $\check{R}$. Each of these two components admits a standard Belov-Moore action. Ignoring the zero modes, the Lagrangian subspaces of $\Omega^{2\ell+1}(M)$ entering the definition of these two Belov-Moore actions are the two most obvious ones that can be constructed for an arbitrary Riemannian manifold $M$, respectively the space of exact forms, and the space of forms that are co-exact with respect to a reference metric $g_0$. For the first choice of Lagrangian, $\hat{R}$, being exact, is pure gauge. So in this case the theory turns out not to have any non-zero modes. For the second choice of Lagrangian, there is no extra gauge symmetry and the non-zero modes of $\check{R}$ are all the exact forms. These are the same off shell degrees of freedom as an ordinary abelian gauge field, and therefore twice as many as we would expect for a self-dual field. \footnote{Quite interestingly, a naive attempt at describing the stringy degrees of freedom of the (2,0) theories in six dimensions ends up with the same overcounting of the degrees of freedom, see \cite{Seiberg:1997ax} on page 18.} It is only by creating a hybrid of these two theories and interpreting it as describing a pair of self-dual fields that we have been able to obtain a consistent quantum partition function.

The non-Gaussian features of the theory of a single self-dual field are apparent in its partition function. The latter should be a square root of the expression \eqref{EqFullPartF} we found for a pair of self-dual fields. This means that it involves various functional determinants, like for instance the determinant of the Hodge Laplacian on $\Omega_{\rm ex}^{2\ell+1}$, to the power $\frac{1}{4}$. This is very surprising if we imagine that the self-dual field is described by a Gaussian action, as a path integral over such an action would produce square roots of functional determinants. Note that the power $\frac{1}{4}$ is genuine, in the sense that there is no simple way to reexpress the determinant of the Laplacian on $\Omega^{2\ell+1}_{\rm ex}$ as the square of another functional determinant. 

The previous paragraph can be reformulated in more physical terms as follows. Let us try to figure out what the off shell degrees of freedom of a self-dual field on a Riemannian manifold should be. Heuristically, a self-dual field should admit half of the degrees of freedom of an ordinary $2\ell$-form abelian gauge field. The off shell degrees of freedom of the latter can be described by exact $2\ell+1$ forms on $M$. The abelian gauge field is a Gaussian theory, that is an infinite collection of non-interacting Gaussian modes whose masses are determined by the eigenvalues of the Hodge Laplacian on $\Omega^{2\ell+1}_{\rm ex}$ (ignoring the ghosts). The effect of these modes appear in the quantum partition function of the theory through a factor proportional to the square root of the determinant of the Laplacian. If the self-dual field was a Gaussian theory, it would mean that there exists a way to split up these modes into a disjoint union of two sets with the same collection of eigenvalues of the Laplacian. This is simply not true on an arbitrary Riemannian manifold.

The previous arguments are admittedly very heuristic. As we have seen, the fact that the partition function is a section of a line bundle makes the issue rather subtle. Also, one could in principle imagine that there exists some complicated second order differential operator such that its determinant coincides with the square root of the determinant of the Laplacian on $\Omega^{2\ell+1}_{\rm ex}$. But we feel that this is a very unlikely possibility.

To our knowledge, the issue of defining the quantum self-dual field on an arbitrary Riemannian manifold by means of an action has not been raised in the literature. Certainly, many actions that reproduce the correct classical equations of motion have been constructed (for a sample, see for instance \cite{Henneaux:1988gg, McClain:1990sx, Pasti:1995tn, Pasti:1996vs, Devecchi:1996cp, Belov:2006jd} and references therein). The Henneaux-Teitelboim is known to reproduce the correct local anomaly of the self-dual field \cite{PhysRevLett.63.728}, but it can be defined only when $M$ admits a non-vanishing vector field, what is in general not the case. The covariant action of Pasti-Sorokin-Tonin \cite{Pasti:1995tn, Pasti:1996vs} requires a non-vanishing vector field as well. The path integral of a manifestly covariant action involving an infinite number of auxiliary fields has been investigated in \cite{Devecchi:1996cp}, but the computations have not been performed in a manifestly covariant gauge, and therefore would probably be difficult to carry over to an arbitrary Riemannian manifold.

In conclusion, it would be interesting to try to study the path integral of the numerous existing actions for the self-dual field, in order to understand this issue better. In doing so, one should be aware that it is not because a classical action reproduces the correct equations of motion of the self-dual field that it will necessarily yield the correct quantum partition function upon path integration. The latter depends on the whole (Euclidean) configuration space of the theory (the kinematics), not only on the configurations solving the equations of motion. Therefore one has to check carefully that the off-shell degrees of freedom are really those expected for a self-dual field. As we have pointed out above, on a generic Riemannian manifold, it is not even clear what these degrees of freedom should be. In our opinion, this indicates that the theory is not of the Gaussian type. At any rate, the abelian self-dual field theory on generic Riemannian manifolds seems to be a rather exotic quantum field theory, and we feel that clarifying this issue might also provide some useful intuition about the mysterious (2,0) superconformal field theories. \\

As far as we are concerned, at the price of extra efforts we will still be able to determine the anomaly bundle of a single self-dual field from the knowledge of the partition function for a pair of self-dual fields. But before starting to study the anomaly bundle, we have to review a bit of modular geometry.

\section{Relevant topics in modular geometry}

\label{SecModGeom}

In this section, we review some material that will be necessary in order to describe the topological anomaly of the self-dual field theory. We have done our best to be self-contained, what inevitably makes this material a bit dense. We try to emphasize the enlightening point of view picturing modular forms as pull-backs of sections of line bundles over modular varieties, which is rarely presented very explicitly in the literature. Sections \ref{SecMGBas} to \ref{SecFacAut} cover standard material \cite{MR2062673}: the definition of modular varieties, of Siegel theta constants and of factors of automorphy. Section \ref{SecMGPicG} contains some recent \cite{Sato01012010, 2009arXiv0908.0555P} or less standard \cite{springerlink:10.1007/BF01231183} material, providing a computation of the topological Picard groups of two modular varieties of interest to us. Section \ref{SecMGBun} uses factors of automorphy in order to compare various bundles over modular varieties, and is in essence contained in \cite{Igusa1964}.

\subsection{Basics}

\label{SecMGBas}

We start with a real $2n$-dimensional symplectic vector space $V$ (in our case $V = H^{2\ell+1}(M,\mathbbm{R})$) with symplectic form $\omega$. Recall that a complex structure $\ast$ on $V$ is said to be compatible with $\omega$ if the bilinear form $\omega(\bullet,\ast \bullet)$ is positive definite.

The integral symplectic group ${\rm Sp}(2n,\mathbbm{Z})$ is the group formed by $2n \times 2n$ integer-valued matrices
\be
\label{EqDecBlockGamma}
\gamma = \begin{pmatrix} A \!& B \\ C \!& D \end{pmatrix} \;,
\ee
satisfying $A^tB = B^tA$, $C^tD = D^tC$ and $AD^t - BC^t = \mathbbm{1}_n$. The superscript $t$ denotes the matrix transposition. ${\rm Sp}(2n,\mathbbm{Z})$ has two important infinite families of finite index normal subgroups. $\Gamma_{2n}^{(k)}$, the level $k$ principal congruence subgroup, is defined as the kernel of the map ${\rm Sp}(2n,\mathbbm{Z}) \rightarrow {\rm Sp}(2n,\mathbbm{Z}_k)$ taking an integral symplectic matrix to its reduction modulo $k$. Note that $\Gamma_{2n}^{(1)} = {\rm Sp}(2n,\mathbbm{Z})$. $\Gamma_{2n}^{(k,2k)}$ is the subgroup of $\Gamma_{2n}^{(k)}$ satisfying the extra conditions $(AB^t)_0 = (CD^t)_0 = 0 \; {\rm mod} \; 2k$. The notation $(M)_0$ denotes the vector formed by the diagonal entries of the matrix $M$. 

We endow $V$ with a Lagrangian decomposition $V = V^0_1 \oplus V^0_2$. We keep the notation of Section \ref{SecRestHarm}: we write $\alpha_i$ and $\beta^i$ for the vectors of a compatible Darboux basis and denote by $a^i$ and $b_i$ the dual coordinates. $\Lambda$ will denote the lattice generated by $\{\alpha_i, \beta^i\}$. Given $v = a^i \alpha_i + b_i \beta^i$, we define the action of ${\rm Sp}(2n,\mathbbm{Z})$ to be
\be
\gamma \cdot \alpha_i = A_i^{\;j} \alpha_j + B_{ij} \beta^j \;, \quad \gamma \cdot \beta^i = C^{ij} \alpha_j + D^i_{\;j} \beta^j \;.
\ee

The space of all compatible affine complex structures on $V$ is parameterized by the Siegel upper half-plane $\mathcal{C}_n$, that is all complex symmetric $n \times n$ matrices $\tau$ with positive definite imaginary part. As in Section \ref{SecRestHarm}, $\tau$ defines the holomorphic coordinates $\{z_i\}$ on $V$ through
\be
\label{EqDef2HolCordV}
z_i = a^i\tau_{ij} + b_i   \;.
\ee
Under the action of ${\rm Sp}(2n,\mathbbm{Z})$, we have
\be
\label{EqModTransZ}
z_i \mapsto a^k (A_k^{\;j} \tau_{ji} + B_{ki}) + b_k (C^{kj} \tau_{ji} + D^k_{\;i}) \;.
\ee
Performing a change of basis $z \mapsto z(C \tau + D)^{-1}$ to go back to coordinates of the form \eqref{EqDef2HolCordV}, we see that ${\rm Sp}(2n,\mathbbm{Z})$ acts on $\mathcal{C}_n$ as follows:
\be
\label{EqActHP}
\qquad \gamma.\tau = (A\tau + B)(C\tau + D)^{-1} \;. 
\ee
This action generalizes the famous action of ${\rm SL}(2,\mathbbm{Z})$ on the complex upper half-plane. We will denote the quotients of $\mathcal{C}_n$ by the action of $\Gamma_{2n}^{(k)}$ and $\Gamma_{2n}^{(k,2k)}$ respectively by $\mathcal{T}_n^{(k)}$ and $\mathcal{T}_n^{(k,2k)}$. These quotients are smooth, except for orbifold singularities if $k \leq 2$. $\mathcal{C}_n$ is contractible, therefore the fundamental groups of $\mathcal{T}_n^{(k)}$ and $\mathcal{T}_n^{(k,2k)}$ are given respectively by $\Gamma_{2n}^{(k)}$ and $\Gamma_{2n}^{(k,2k)}$. 

As $n$ will be fixed during all our discussion, we will drop the subscripts related to $n$ in the following. Also, $\Gamma$ and $\Gamma'$ will denote finite index subgroups of ${\rm Sp}(2n,\mathbbm{Z})$ and $\mathcal{T}$ and $\mathcal{T}'$ the associated quotient of $\mathcal{C}$. Using this notation, whenever we have $\Gamma \subset \Gamma'$, then $\mathcal{T}$ is a covering of $\mathcal{T}'$. The degree of the covering is equal to the index of $\Gamma$ in $\Gamma'$ and is therefore finite.

We define the group of line bundles on $\mathcal{T}$ as the group of oriented $\Gamma$-equivariant complex line bundles on $\mathcal{C}$ (with the tensor product as the group operation). This group is called the topological Picard group of $\mathcal{T}$, denoted ${\rm Pic}(\mathcal{T})$. There is an injection ${\rm Pic}(\mathcal{T}) \hookrightarrow H^2_\Gamma(\mathcal{C},\mathbbm{Z})$ assigning to a line bundle its first Chern class (see Section 2.2 of \cite{2009arXiv0908.0555P} for more details and references). If $\Gamma \subset \Gamma'$, then any $\Gamma'$-equivariant bundle on $\mathcal{C}$ is also a $\Gamma$-equivariant bundle and we have a map ${\rm Pic}(\mathcal{T'}) \rightarrow {\rm Pic}(\mathcal{T})$. This map coincides with the pull-back with respect to the covering $\mathcal{T} \rightarrow \mathcal{T'}$.

The Hodge bundle over $\mathcal{T}$ is the rank $n$ bundle whose fiber over $\tau \in \mathcal{T}$ is the holomorphic tangent space of $V$, the holomorphicity condition referring to the complex structure corresponding to $\tau$. We will denote its determinant bundle by $\mathscr{K}$. $\mathscr{K}$ is a very useful line bundle to consider, because it is defined on $\mathcal{C}/\Gamma$ for any finite-index subgroup $\Gamma$.

\subsection{Siegel theta constants}

\label{SecSiegThet}

The classical Siegel theta functions are holomorphic functions on $V \times \mathcal{C}$, defined by 
\be
\label{EqDefThetaFunc}
\theta^\eta(Z, \tau) = \sum_{R \in \Lambda_1 + \eta_1} \exp \big(\pi i r^i \tau_{ij} r^j + 2\pi i (z_i + \eta^2_i)r^i \big) \;,
\ee
where we used the notation of equation \eqref{EqFPartZeroModes}. They depend on a half-integral characteristic $\eta \in \frac{1}{2}\Lambda$. The components of $\eta$ on the components of a Lagrangian decomposition $V = V^0_1 \oplus V^0_2$ are written $\eta_1$ and $\eta^2$, respectively. A characteristic $\eta$ is called odd or even, depending on the parity of $4\sum_i(\eta_1^i \eta_1^i + \eta^2_i \eta^2_i)$. Characteristics differing by an integral vector give rise to the same theta function up to a constant factor:
\be
\label{EqTransThetaCharShift}
\theta^{\eta+\lambda}(Z, \tau) = \exp(2 \pi i \eta_1^i \lambda^2_i) \; \theta^{\eta}(Z, \tau) \;,
\ee
for $\lambda \in \Lambda$.

The action of ${\rm Sp}(2n,\mathbbm{Z})$ on $V \times \mathcal{C}$ described in the previous section acts on the theta functions, according to the theta transformation formula:
\be
\label{EqThetaTrans}
\theta^{\gamma \ast \eta}(Z(C\tau + D)^{-1}, \gamma.\tau) = \kappa_\eta(\gamma) \det(C\tau + D)^{\frac{1}{2}} \exp(\pi i Z (C\tau + D)^{-1} C Z^t) \theta^\eta(Z, \tau) \;,
\ee
where $\kappa_\eta(\gamma)$ is a constant depending only on $\gamma$ and $\eta$. $\kappa_\eta(\gamma)$ also depends on the choice of a branch for the square root $\det(C\tau + D)^{\frac{1}{2}}$ that we make once and for all. Once such a choice is made, $\kappa_\eta(\gamma)$ can be computed explicitly \cite{Stark1982, Styer1984}. The induced action of ${\rm Sp}(2n,\mathbbm{Z})$ on the characteristics is given explicitly by
\be
\label{EqTransEta}
\gamma \ast \eta = (\gamma^t)^{-1} \cdot \eta + \frac{1}{2}((C D^t)_0)^i \alpha_i + \frac{1}{2}((A B^t)_0)_i \beta^i \;.
\ee
Note that this is an affine action, not a linear one, and that it preserves the parity of the characteristic. From their definitions, we can immediately see that the group $\Gamma^{(2)}$ preserves all characteristics modulo $\Lambda$ and that $\Gamma^{(1,2)}$ leaves the characteristic $\eta = 0$ fixed modulo $\Lambda$.

Theta constants are the holomorphic functions on $\mathcal{C}$ obtained by the evaluation of the theta functions at $Z = 0$. We will simply write them $\theta^\eta(\tau)$. Their transformation formula follows from \eqref{EqThetaTrans}:
\be
\label{EqThetaConstTrans}
\theta^{\gamma \ast \eta}(\gamma.\tau) = \kappa_\eta(\gamma) \det(C\tau + D)^{\frac{1}{2}} \theta^\eta(\tau) \;.
\ee
Theta constants with odd characteristic vanish.

\subsection{Factors of automorphy}

\label{SecFacAut}

The fact that $\mathcal{C}$ is contractible means that any bundle on $\mathcal{T}$ has to pull back to a trivial bundle on $\mathcal{C}$. Holomorphic line bundles on a quotient space that pull back to the trivial bundle can be described very effectively by means of factors of automorphy.

A factor of automorphy is a 1-cocycle on the fundamental group of $\mathcal{T}$ with value in the multiplicative group of non-vanishing holomorphic functions on $\mathcal{C}$. The group of such cocycles is written $Z^1\big(\pi_1(\mathcal{T}),H^0(\mathscr{O}^\ast_\mathcal{C})\big)$. By the previous definition, a factor of automorphy $\xi$ associates to each element $\gamma \in \pi_1(\mathcal{T})$ a non-vanishing holomorphic function $\xi_\gamma(\tau)$ on $\mathcal{C}$, $\tau \in \mathcal{C}$. The cocycle condition reads
\be
\label{EqCocCondFactAut}
\xi_{\gamma'\gamma}(\tau) = \xi_{\gamma'}(\gamma.\tau) \xi_{\gamma}(\tau) \;,
\ee
where $\gamma.\tau$ denotes the action of $\pi_1(\mathcal{T})$ on the universal covering $\mathcal{C}$. 

To construct a holomorphic line bundle on $\mathcal{T}$ out of a factor of automorphy, we take the quotient of $\mathcal{C} \times \mathbbm{C}$ by the following action of $\pi_1(\mathcal{T})$:
\be
\gamma.(\tau, w) = (\gamma.\tau, \xi_\gamma(\tau)w) \;.
\ee
We see that the factor of automorphy describes the change of trivialization of the pulled-back line bundle on $\mathcal{C}$ between patches related by the action of $\pi_1(\mathcal{T})$. The cocycle condition on the factor of automorphy ensures that these changes of trivializations give rise to a well-defined line bundle. An immediate consequence is that the factor of automorphy associated with the tensor product of two line bundles is given by the product of the factors of automorphy of the factors. 

Different factors of automorphy do not necessarily give rise to different line bundles. Given any non-vanishing holomorphic function $f$ on $\mathcal{C}$ and a factor of automorphy $\xi$, define
\be
\label{EqEquivFactAut}
\xi'_\gamma(\tau) = \frac{f(\gamma.\tau)}{f(\tau)} \xi_\gamma(\tau)\;.
\ee
Then the factors $\xi$ and $\xi'$ describe the same holomorphic line bundle. Indeed, if we perform the change of trivialization defined by $f$, the transition functions between the patches will transform according to \eqref{EqEquivFactAut}. The 1-cocycle $\frac{f(\gamma.\tau)}{f(\tau)} \in Z^1\big(\pi_1(\mathcal{T}),H^0(\mathscr{O}^\ast_\mathcal{C})\big)$ is the differential of the 0-cochain $f(\tau)$ with respect to the group cohomology differential. Therefore \eqref{EqEquivFactAut} tells us that any two factors of automorphy differing by an exact factor give rise to the same holomorphic line bundle. It can be shown that indeed, the group of holomorphic line bundles on $\mathcal{T}$ is isomorphic to the cohomology group $H^1\big(\pi_1(\mathcal{T}),H^0(\mathscr{O}^\ast_\mathcal{C})\big)$ (see the appendix B of \cite{MR2062673} for a proof). 

Sections of holomorphic line bundles on $\mathcal{T}$ can be nicely described in terms of factors of automorphy. A section, when pulled back on $\mathcal{C}$, will give rise to a holomorphic function $s$ satisfying the functional equation
\be
\label{EqRelFactAutSec}
s(\gamma.\tau) = \xi_\gamma(\tau)s(\tau) \;.
\ee
This is a consequence of the fact that factors of automorphy describe changes of trivialization between patches on $\mathcal{C}$. As a nice consistency check, note that if $s$ does not vanish anywhere, then $\xi_\gamma(\tau)$ can be expressed as an exact cocycle $s(\gamma.\tau)/s(\tau)$, and the corresponding line bundle should be trivial. But this is of course true, because a non-vanishing section of a line bundle necessarily trivializes it globally.

A consequence of the interpretation of the factors of automorphy as transition functions is that a bundle is flat if and only if it admits a factor of automorphy that is constant in $\tau$ for all $\gamma$. The latter is nothing but a character of $\pi_1(\mathcal{T})$, parameterizing the holonomies of the flat bundle along the homotopically non-trivial loops in $\mathcal{T}$.

We would like to compute two examples of factors of automorphy to conclude this section. Our first example is the determinant $\mathscr{K}$ of the Hodge bundle defined in Section \ref{SecMGBas}. $\mathscr{K}$ admits a natural section:
\be
\label{EqSecK}
s = \zeta^1 \wedge ... \wedge \zeta^{n} \;,
\ee
where $\zeta^i$ are the holomorphic vectors defined in \eqref{EqDefZeta}.
Under a modular transformation $\gamma \in {\rm Sp}(2n,\mathbbm{Z})$, \eqref{EqModTransZ} implies that $\zeta_i \rightarrow \zeta^j (C\tau + D)_{j}^{\;i}$. From this, we deduce the transformation of $s$:
\be
s(\gamma.\tau) = \det(C\tau +D) s(\tau) \;.
\ee 
Comparing with \eqref{EqRelFactAutSec}, we see that we just determined a factor of automorphy for the determinant of the Hodge bundle. It reads explicitly:
\be
\label{EqFactAutDetHodge}
\xi^{\mathscr{K}}_\gamma(\tau) = \det(C\tau + D) \;.
\ee
It can be checked that that $\xi^{\mathscr{K}}$ does satisfy the cocycle condition \eqref{EqCocCondFactAut}. This little computation also allows one to show that Siegel modular forms of weight $k$ are pull-backs on $\mathcal{C}$ of sections of $\mathscr{K}^k$. Indeed, their definition requires the corresponding factor of automorphy to be $\det(C\tau + D)^k$. 

Our second example is the theta constant with characteristic $0$. Equation \eqref{EqTransThetaCharShift} shows that integral shifts of the $0$ characteristic leaves the corresponding theta constant invariant. From the transformation formula \eqref{EqThetaConstTrans}, we can read off the factor of automorphy of the bundle $\mathscr{C}^0$ admitting it as a section:
\be
\label{EqFactAutTheta0}
\xi^{\mathscr{C}^0}_\gamma(\tau) = \kappa_0(\gamma) \det(C\tau + D)^{\frac{1}{2}} \;.
\ee
The bundle $(\mathscr{C}^0)^2 \otimes \mathscr{K}^{-1}$ admits the factor of automorphy $\xi_\gamma = (\kappa_0(\gamma))^2$, which is independent of $\tau$. We deduce that this bundle is flat and that modulo torsion, $(\mathscr{C}^0)^2 = \mathscr{K}$.

The factor \eqref{EqFactAutTheta0} describes a transition function for the bundle $\mathscr{C}^0$ between two patches related by the transformation $\gamma$. But as \eqref{EqFactAutTheta0} has not norm $1$ in general, the transition function is not unitary, meaning that the trivialization in which the theta function takes its standard form is not unitary. To make it unitary, we have to perform a change of trivialization by multiplying the theta function by the non-vanishing real function $(\det h)^{1/4}$. Indeed, writing $\gamma.h$ for $\gamma.\tau - \overline{\gamma.\tau}$, we have 
\be
\det \gamma.h = \overline{\det(C\tau + D)}^{-1} \det h \det(C\tau + D)^{-1} \;,
\ee
so that $(\det h)^{1/4} \theta^{\eta = 0}$ transforms with a multiplier of modulus 1. Therefore, the norm of the section $s$ of $\mathscr{C}^0$ pulling back to the theta function in the standard trivialization reads
\be
(s,s) = \sqrt{h} \,|\theta^{\eta = 0}|^2 \;.
\ee
Comparing with \eqref{EqHermStrucKalPot}, we see that the Kähler potential of $\mathscr{C}^0$ is given by $-\frac{1}{2} \ln \det h$.

\subsection{Topological Picard groups}

\label{SecMGPicG}

In this section, we report on some result about the topological Picard groups of the quotients $\mathcal{T}^{(2)}$ and $\mathcal{T}^{(1,2)}$ \cite{2009arXiv0908.0555P, Sato01012010, springerlink:10.1007/BF01231183}. Our interest in these particular quotients stems from the fact that $\Gamma^{(2)}$ is the subgroup of ${\rm Sp}(2n,\mathbbm{Z})$ that keeps all the theta characteristics fixed, while $\Gamma^{(1,2)}$, the so-called theta group, is the group leaving the zero characteristic fixed.

\subsubsection*{The quotient by the level 2 principal congruence group}

This is the space $\mathcal{T}^{(2)}$ and we denote its topological Picard group by ${\rm Pic}(\mathcal{T}^{(2)})$. Specializing the results of \cite{2009arXiv0908.0555P} and provided $n > 2$, we have a short exact sequence
\be
\label{EqSeqPic}
0 \rightarrow {\rm Hom}\big(\Gamma_{\rm ab}^{(2)}, \mathbbm{Q}/\mathbbm{Z}\big) \rightarrow {\rm Pic}(\mathcal{T}^{(2)}) \rightarrow \mathbbm{Z} \rightarrow 0 \;,
\ee
where $\Gamma_{\rm ab}^{(2)}$ denotes the abelianization of $\Gamma^{(2)}$. The abelianization of a group $\Gamma$ is the quotient of $\Gamma$ by its commutator subgroup, and any character of $\Gamma$ factorizes through its abelianization. We saw in previous sections that flat bundles over $\mathcal{T}^{(2)}$ are classified by the characters of $\pi_1(\mathcal{T}^{(2)})$. We also saw that $\pi_1(\mathcal{T}^{(2)}) \simeq \Gamma^{(2)}$, because $\mathcal{T}^{(2)}$ is the quotient of a contractible space by $\Gamma^{(2)}$. Therefore the characters ${\rm Hom}(\Gamma_{\rm ab}^{(2)}, \mathbbm{Q}/\mathbbm{Z})$ classify the flat line bundles on ${\rm Pic}(\mathcal{T}^{(2)})$. 

The projection on $\mathbbm{Z}$ corresponds to projecting the integral first Chern class onto its image in real cohomology. Moreover, it can be shown that $\mathscr{K}$ projects on $2 \in \mathbbm{Z}$. From the factor of automorphy \eqref{EqFactAutDetHodge} of $\mathscr{K}$, we see that any bundle that admits a factor of automorphy of the form $c \det(C\tau + D)^{k/2}$, $c$ independent of $\tau$, projects on $k \in \mathbbm{Z}$.

$\Gamma_{\rm ab}^{(2)}$ was computed explicitly in the Section 2 of \cite{Sato01012010}: 
\be
\Gamma_{\rm ab}^{(2)} = (\mathbbm{Z}_2)^{2n^2-n} \times (\mathbbm{Z}_4)^{2n} \;. 
\ee
The projection of $\Gamma^{(2)}$ on its abelianization can be described explicitly. Let us decompose an element $\gamma \in \Gamma^{(2)}$ into $n \times n$ blocks as follows:
\be
\label{EqDecBlockGamma2}
\gamma - \mathbbm{1} = 2 \begin{pmatrix} \tilde{A}(\gamma) \!& \tilde{B}(\gamma) \\ \tilde{C}(\gamma) \!& \tilde{D}(\gamma) \end{pmatrix} \;.
\ee
There is a homomorphism $m : \Gamma^{(2)} \rightarrow (\mathbbm{Z}_2)^{2n^2-n} \times (\mathbbm{Z}_4)^{2n}$ given by
\begin{align}
\label{EquMapAb}
m(\gamma) =  \big( &\, \{\tilde{A}_{ij}(\gamma)\}_{1 \leq i \leq n, 1 \leq j \leq n} \; {\rm mod} \; 2, \notag \\
&\, \{\tilde{B}_{ij}(\gamma)\}_{1 \leq i < j \leq n} \; {\rm mod} \; 2, \notag \\
&\,\{\tilde{C}_{ij}(\gamma)\}_{1 \leq i < j \leq n} \; {\rm mod} \; 2,  \\
&\,\{\tilde{B}_{ii}(\gamma)\}_{1 \leq i \leq n} \; {\rm mod} \; 4, \notag \\
&\,\{\tilde{C}_{ii}(\gamma)\}_{1 \leq i \leq n} \; {\rm mod} \; 4 \big) \notag \;.
\end{align}
The components of the map $m$ form a basis for the (additive) characters of $\Gamma^{(2)}$. We will call these components the elementary characters of $\Gamma^{(2)}$. Equivalently, they form a system of generators for the group of flat line bundles on $\mathcal{T}^{(2)}$.

It will be useful for us to consider a system of generators of $\Gamma^{(2)}$ that is in a sense dual to the basis of additive characters presented above. 
We define the following elements of ${\rm Sp}(2n,\mathbbm{Z})$ \cite{Igusa1964}:
\begin{itemize}
	\item $\alpha^{(ij)}$, the $2n \times 2n$ identity matrix with the entry $(i,j)$ replaced by $2$ and the entry $(n+j, n+i)$ replaced by $-2$, for $1 \leq i \leq n$, $1 \leq j \leq n$ and $i \neq j$;
	\item $\alpha^{(ii)}$, the $2n \times 2n$ identity matrix with the entries $(i,i)$ and $(n+i, n+i)$ replaced by $-1$, for $1 \leq i \leq n$;
	\item $\beta^{(ij)}$, the $2n \times 2n$ identity matrix with the entries $(i,n+j)$ and $(j, n+i)$ replaced by $2$ , for $1 \leq i \leq n$, $1 \leq j \leq n$ and $i < j$;
	\item $\gamma^{(ij)} := (\beta^{(ij)})^t$;
	\item $\beta^{(ii)}$, the $2n \times 2n$ identity matrix with the entry $(i,n+i)$ replaced by $2$, for $1 \leq i \leq n$;
	\item $\gamma^{(ii)} := (\beta^{(ii)})^t$.
\end{itemize}
It is straightforward to check that these matrices actually belong to $\Gamma^{(2)}$ and that they generate it. We will call these generators the elementary generators. For each elementary additive character forming the components of the map $m$, there is one of the elementary generators that is mapped to $1$ by this character and to $0$ by all the other characters. This is the duality property that we hinted at above. Practically, given any character of $\Gamma^{(2)}$, its evaluation on the elementary generators allows one to express it in terms of the elementary additive characters. 

Finally, let us stress that the restriction to $n > 2$ is very important. We are not aware of a computation of the Picard groups of $\mathcal{T}^{(2)}$ for $n = 1$ or $n = 2$. However the topological Picard groups of the quotient $\mathcal{T}^{(1)} := \mathcal{C}/{\rm Sp}(2n,\mathbbm{Z})$ show that these cases are quite different: ${\rm Pic}(\mathcal{T}^{(1)})$ is isomorphic to $\mathbbm{Z}_{12}$ for $n = 1$, to $\mathbbm{Z}_{10}$ for $n = 2$ and to $\mathbbm{Z}$ for $n > 2$ (see \cite{MR1795866}, Section 17).

\subsubsection*{The quotient by the theta group}

We now turn to $\mathcal{T}^{(1,2)}$. We will be essentially interested in the group of flat line bundles, or equivalently in (the character group of) the abelianization $\Gamma^{(1,2)}_{\rm ab}$ of $\Gamma^{(1,2)}$. The latter was computed in \cite{springerlink:10.1007/BF01231183}:
\be
\Gamma^{(1,2)}_{\rm ab} = \mathbbm{Z}_4 \;.
\ee
To describe the quotient map $\Gamma^{(1,2)} \rightarrow \Gamma^{(1,2)}_{\rm ab}$, consider the elements of ${\rm Sp}(2n,\mathbbm{Z})$ defined by
\be
\gamma_v(w) = w + \omega(v,w)v
\ee
for $v \in \Lambda \subset V$. Such a symplectic transformation is called a transvection. Consider moreover the $\mathbbm{Z}_2$-valued bilinear form $Q$ sending $w = (w_1, w^2)$ to $\sum_i w^i_1 w^2_i$ modulo 2. If $Q(v) = 1$, the transvection is called anisotropic. In \cite{springerlink:10.1007/BF01231183}, it is shown that $\Gamma^{(1,2)}$ is generated by the conjugacy class of an anisotropic transvection. As a result, we can get an explicit map $\Gamma^{(1,2)} \rightarrow \Gamma^{(1,2)}_{\rm ab}$ by sending this anisotropic transvection to $1 \in \mathbbm{Z}_4$.

Recall that the transformation formula for theta constant with zero characteristic is involving a certain set of constants $\kappa_0(\gamma)$. At the end of Section \ref{SecFacAut}, we showed that $(\kappa_0(\gamma))^2$ is the factor of automorphy of a flat line bundle $(\mathscr{C}^0)^2 \otimes \mathscr{K}^{-1}$. This bundle is well-defined on $\mathcal{T}^{(1,2)}$, so $(\kappa_0(\gamma))^2$ should be a character of $\Gamma^{(1,2)}$. \footnote{Note that while $(\kappa_0(\gamma))^2$ is a character, $\kappa_0(\gamma)$ is not. The latter can be interpreted as a character by considering the double cover of the symplectic group, the metaplectic group. See \cite{springerlink:10.1007/BF01231183}. This is not needed for our purpose, however.} It was proven in \cite{springerlink:10.1007/BF01231183} that this character takes the value $i$ on a certain class of anisotropic transvections. Therefore it is a generator of the group of characters of $\Gamma^{(1,2)}$. More explicitly, the following formula holds \cite{springerlink:10.1007/BF01231183}:
\be
\label{EqCharKappa0}
(\kappa_0(\gamma))^2 = i^{-r} \epsilon(\det E(\gamma)) \;.
\ee
$r$ is the rank of the block $C$ in $\gamma$ after reduction modulo 2. $E$ is a matrix obtained from $C$ by selecting $r$ linearly independent rows and replacing the other ones by the corresponding rows of the block $A$ of $\gamma$. Finally, $\epsilon(m) = 0$ if $m$ is even, $1$ if $m = 1$ modulo $4$ and $-1$ if $m = 3$ modulo $4$.

As $\Gamma^{(2)} \subset \Gamma^{(1,2)}$, we have an inclusion of character groups 
\be
{\rm Hom}\big(\Gamma_{\rm ab}^{(1,2)}, \mathbbm{Q}/\mathbbm{Z}\big) \rightarrow {\rm Hom}\big(\Gamma_{\rm ab}^{(2)}, \mathbbm{Q}/\mathbbm{Z}\big) \;.
\ee
The image of this inclusion gives us the flat bundles on $\mathcal{T}^{(2)}$ that are pull-backs of flat line bundles on $\mathcal{T}^{(1,2)}$. The explicit formula \eqref{EqCharKappa0} allows us to compute this pull-back map. 

To this end, we consider the set of generators of $\Gamma^{(2)}$ dual to the basis of characters provided by \eqref{EquMapAb} and described in the previous section. Recall that for each choice of character among $\tilde{A}_{ij}$, $\tilde{B}_{ij}$ and $\tilde{C}_{ij}$, we found an element of $\Gamma^{(2)}$ such that the selected character has value $1$ on this element and all other character vanish. \footnote{We chose an additive notation for these characters.} Given such an element $\gamma$ of $\Gamma^{(2)}$, we compute $(\kappa_0(\gamma))^2$ by means of \eqref{EqCharKappa0}, and we obtain a formula for the additive character corresponding to $(\kappa_0(\gamma))^2$ in terms of a sum of the elementary characters $\tilde{A}_{ij}$, $\tilde{B}_{ij}$ and $\tilde{C}_{ij}$. In multiplicative notation, we obtain
\be
\label{EqCharKappa0PB}
(\kappa_0(\gamma))^2|_{\Gamma^{(2)}} = \exp\left(\pi i \sum_{i} \tilde{A}_{ii}(\gamma)\right) \;.
\ee

Let us summarize. The flat line bundle $(\mathscr{C}^0)^2 \otimes \mathscr{K}^{-1}$ generates the group $\mathbbm{Z}_4$ of flat line bundles on $\mathcal{T}^{(1,2)}$. The corresponding character is given by \eqref{EqCharKappa0}. We have just shown that this bundle pulls back to the flat line bundle on $\mathcal{T}^{(2)}$ whose character is given by \eqref{EqCharKappa0PB}. Note that while the original line bundle had order 4, the pulled back bundle has order 2. Hence there are only two flat bundles on $\mathcal{T}^{(2)}$ that are pull-backs of bundles on $\mathcal{T}^{(1,2)}$: $(\mathscr{C}^0)^2 \otimes \mathscr{K}^{-1}$ and the trivial bundle. 

This remarkable property will allow us in Section \ref{SecAnSingSDF} to identify the anomaly bundle of a single self-dual field from the knowledge of the anomaly bundle of a pair of self-dual fields, up to a very mild ambiguity.

\subsection{Bundles of interest}

\label{SecMGBun}

We now complete our computations of the factors of automorphy of $\mathscr{C}^\eta$ and $\mathscr{K}$ and see how these bundles fit in the topological Picard groups computed in the previous section.

\subsubsection*{Factors of automorphy}

Given an element $\gamma$ of $\Gamma^{(2)}$, let $\lambda(\gamma,\eta) = \gamma \ast \eta - \eta \in \Lambda$. Using the transformation formulas \eqref{EqTransThetaCharShift} and \eqref{EqThetaConstTrans} for classical theta constants, we find the following factor of automorphy for the bundle $\mathscr{C}^\eta$:
\be
\label{EqFactAutCEta}
\xi^{\mathscr{C}^\eta}_\gamma(\tau) = \exp \big(2\pi i \eta^i_1 \lambda^2_i(\gamma,\eta)\big) \, \kappa_\eta(\gamma) \det(C\tau + D)^{\frac{1}{2}}\;,
\ee
The dependence of $\kappa_\eta(\gamma)$ on $\eta$ can be made explicit:
\be
\kappa_\eta(\gamma) = \kappa_0(\gamma) \exp(\pi i k(\gamma, \eta)) \;,
\ee
where
\be
\label{EqDefK}
k(\gamma, \eta) = (D\eta_1 - C \eta^2)^i(-B\eta_1 + A\eta^2 + (AB^t)_0)_i - \eta_1^i \eta^2_i \;.
\ee
As was already mentioned, $\kappa_0$ depends on the choice of branch for $\det(C\tau + D)^{\frac{1}{2}}$ that we fixed throughout the whole discussion and can be computed explicitly \cite{Stark1982, Styer1984}. However we will not need its explicit expression.

The factor of automorphy for $\mathscr{K}$ has been computed in equation \eqref{EqFactAutDetHodge}.

\subsubsection*{Comparing bundles}

Given two line bundles $\mathscr{L}$ and $\mathscr{L}'$ on $\mathcal{T}^{(2)}$ with the same first Chern class modulo torsion, the product $\mathscr{L}^{-1} \otimes \mathscr{L}'$ is a flat bundle. By ``comparing'' two line bundles, we mean computing the character of $\Gamma^{(2)}$ associated with $\mathscr{L}^{-1} \otimes \mathscr{L}'$. Expressing this character in terms of the elementary additive characters $\{\tilde{A}_{ij}\}$, $\{\tilde{B}_{ij}\}$ and $\{\tilde{C}_{ij}\}$ provides a simple mean of checking whether the two bundles are actually isomorphic or not.

First, from the powers of $\det(C\tau + D)$ involved in their factors of automorphy, we see that all the bundles $\mathscr{C}^\eta$ have the same real Chern class and project on $1 \in \mathbbm{Z}$ under the second arrow of \eqref{EqSeqPic}. We will choose $\mathscr{C}^0$ as a reference bundle and compare the bundles $\mathscr{C}^\eta$ to it. From the multiplicative property of factors of automorphy under tensor products, we have 
\be
\label{EqCharCetaCzero}
\xi^{\mathscr{C}^\eta \otimes (\mathscr{C}^0)^{-1}}_\gamma = \exp \big(2\pi i \eta^i_1 \lambda^2_i(\gamma,\eta)\big) \, \exp(\pi i k(\gamma, \eta)) \;.
\ee
To find out if this character is trivial, we express it in terms of the elementary characters \eqref{EquMapAb}. Practically, this amounts to evaluating \eqref{EqCharCetaCzero} on the elementary generators of $\Gamma^{(2)}$ that we defined in Section \ref{SecMGPicG}. \footnote{This computation has essentially been performed by Igusa in his classical paper on theta constants \cite{Igusa1964}, see Theorem 3 there. The apparent discrepancy between his result and \eqref{EqCharCetaCzeroExpl} is due to the fact that his characteristics are valued in $(\mathbbm{Z}_2)^{2n}$.} The result reads:
\be
\label{EqCharCetaCzeroExpl}
\xi^{\mathscr{C}^\eta \otimes (\mathscr{C}^0)^{-1}}_\gamma = \exp \left( \pi i n_1(\gamma,\eta) + \frac{\pi i}{2} n_2(\gamma,\eta) \right) \;,
\ee
with
\begin{align}
n_1(\gamma,\eta)/4 = &\, \sum_{i,j} \tilde{A}_{ij}(\gamma) \eta_1^i \eta^2_j - \sum_{i < j}  \tilde{B}_{ij}(\gamma) \eta_1^i \eta_1^j - \sum_{i < j}  \tilde{C}_{ij}(\gamma) \eta^2_i \eta^2_j \;, \\
n_2(\gamma,\eta)/4 = &\, -\sum_{i} \tilde{B}_{ii}(\gamma) \eta_1^i(\eta_1^i-1) - \sum_{i} \tilde{C}_{ii}(\gamma) (\eta^2_i)^2 \notag \;.
\end{align}
$\xi^{\mathscr{C}^\eta \otimes (\mathscr{C}^0)^{-1}}$ is trivial if and only if $\eta = 0 \;{\rm mod} \; \Lambda$. We immediately deduce that the fiber bundles $\mathscr{C}^\eta$ are all distinct.

Now we would like to compare the square of the theta bundle with characteristic $\eta$ with the determinant of the Hodge bundle, i.e. compute the character associated to the flat bundle $(\mathscr{C}^\eta)^2 \otimes (\mathscr{K})^{-1}$. From the factors of automorphy \eqref{EqFactAutDetHodge} and \eqref{EqFactAutCEta}, we deduce 
\be
\xi^{(\mathscr{C}^\eta)^2 \otimes (\mathscr{K})^{-1}}_\gamma = (\kappa_0(\gamma))^2 \exp(2\pi i k(\gamma, \eta)) \;.
\ee
We computed $(\kappa_0(\gamma))^2$ in terms of elementary characters in \eqref{EqCharKappa0PB}: 
\be
(\kappa_0(\gamma))^2 = \exp (\pi i \sum_j \tilde{A}_{jj})\;,
\ee
from which we deduce
\be
\label{EqCharCeK}
\xi^{(\mathscr{C}^\eta)^2 \otimes (\mathscr{K})^{-1}}_\gamma = \exp \left( \pi i \sum_j \tilde{A}_{jj}(\gamma) - 4\pi i \sum_{i} \tilde{B}_{ii}(\gamma) (\eta_1^i)^2 - 4\pi i\sum_{i} \tilde{C}_{ii}(\gamma) (\eta^2_i)^2 \right) \;.
\ee
This character is non-trivial for all $\eta$. Therefore none of the bundles $\mathscr{C}^\eta$ is a square root of $\mathscr{K}$

\section{The topological anomaly}

\label{SecAnom}

In this section, we identify the anomaly bundle of the self-dual field theory, using results from the previous section.

\subsection{The anomaly bundle of a pair of self-dual fields}

\label{SecClassAn}

Recall that in Section \ref{SecPathInt}, we showed that  the partition function of a pair of self-dual fields could be decomposed into the product of a factor $\mathcal{Z}_{0}$ coming from the sum over topologically distinct configurations of zero modes and a factor $N$ coming from the integration over the non-zero modes:
\be
\mathcal{Z}(Z,\eta) = N \mathcal{Z}_{0}(Z,\eta) \;.
\ee
The anomaly bundle $\mathscr{A}^\eta$ of a pair of self-dual fields decomposes accordingly into a tensor product 
\be
\mathscr{A}^\eta \simeq \mathscr{A}^\eta_0 \otimes \mathscr{A}_+ \;,
\ee 
where $\mathscr{A}_0$ is a bundle admitting $\mathcal{Z}_{0}$ as a section and $\mathscr{A}_+$ is a bundle admitting $N$ as a section.

At the end of Section \ref{SecIntMassive}, we showed that $N$ never vanishes. A line bundle with a non-vanishing section is necessarily trivial, and we learn that $\mathscr{A}_+$ is a topologically trivial bundle. This does not mean that it does not contribute to the local and global anomaly, because it does not come equipped with a natural trivialization. Indeed we saw in Section \ref{SecFPartSecLi} that $N$ does contribute to the Hermitian structure of the anomaly bundle, and hence to its curvature. However, it does not contribute to the topological anomaly and we have $\mathscr{A}^\eta \simeq \mathscr{A}_0^\eta$ as topological bundles.

Now we would like to show that the bundle $\mathscr{A}^\eta$ is actually the pull-back from a bundle defined over the quotient $\mathcal{C}/\Gamma$ of the Siegel upper-half plane by a certain subgroup of ${\rm Sp}(2n,\mathbbm{Z})$. We have a projection $\mathcal{M} \rightarrow \mathcal{C}$ obtained by restricting the Hodge star operator on the space of harmonic $2\ell+1$-forms and identifying the latter with $H^{2\ell+1}$.  Consider first the action of $\mathcal{D}_0$, the connected component of the identity of $\mathcal{D}$. It acts trivially on the cohomology and leaves invariant the action of the Hodge star operator on $H^{2\ell+1}$, so we have a projection $\mathcal{M}/\mathcal{D}_0 \rightarrow \mathcal{C}$. The mapping class group of $M$ is the group of components of the group of diffeomorphisms, defined by ${\rm Mcg} := \mathcal{D}/\mathcal{D}_0$. It acts non-trivially on $H^{2\ell+1}$, but this action preserves the integral symplectic form $\omega$. Therefore it necessarily factorizes through a homomorphism
\be
{\rm Mcg} \rightarrow {\rm Sp}(2n,\mathbbm{Z}) \;.
\ee 
Denote by $\mathcal{D}^{(2)}$ the subgroup of diffeomorphisms that projects in $\Gamma^{(2)}$ under the map 
\be
\mathcal{D} \rightarrow {\rm Mcg} \rightarrow {\rm Sp}(2n,\mathbbm{Z}) \;.
\ee

$\mathcal{Z}_{0}(Z,\eta)$, up to an irrelevant normalization factor, is given by the square of a theta function. If we set the field $Z$ to zero, it is actually a theta constant with characteristic $\eta$. We know that theta constants are sections of bundles $\mathscr{C}^\eta$, defined over $\mathcal{C}/\Gamma^{(2)}$. The bundles $\mathscr{C}^\eta$ pull back to bundles $\tilde{\mathscr{C}}^\eta$ on $\mathcal{M}/\mathcal{D}^{(2)}$. For a given choice of characteristic $\eta$, the anomaly bundle for a pair of self-dual fields is therefore given by
\be
\label{EqAnomBundleId}
\mathscr{A}^\eta \simeq (\tilde{\mathscr{C}}^\eta)^2 \;.
\ee
This is the main result of this section.

Let us make two remarks. 
\begin{itemize}
	\item As the diffeomorphism group acts non-trivially on the characteristic, the partition function can in general only be defined on $\mathcal{M}/\mathcal{D}^{(2)}$, and not on $\mathcal{M}/\mathcal{D}$ \cite{Witten:1996hc}. As is obvious from its definition, $\mathcal{D}^{(2)}$ is the group of diffeomorphisms preserving all characteristics.
	\item Actually, we can slightly refine our statement and consider the group $\mathcal{D}^{(1,2)}$ of diffeomorphisms preserving only the characteristic $\eta = 0$. This is the subgroup of $\mathcal{D}$ mapped to $\Gamma^{(1,2)}$. $(\mathscr{C}^0)^2$ is a well-defined bundle on $\mathcal{T}^{(1,2)} = \mathcal{C}/\Gamma^{(1,2)}$ that pulls back to a bundle on $\mathcal{M}/\mathcal{D}^{(1,2)}$. \eqref{EqAnomBundleId} is true as well when the bundles are interpreted as bundles on this quotient. Of course, for another (even) characteristic $\eta$, the same argument goes through when replacing $\Gamma^{(1,2)}$ by a conjugate subgroup preserving $\eta$.
\end{itemize}

\subsection{The anomaly bundle of a single self-dual field}

\label{SecAnSingSDF}

Let us now consider a single self-dual field. We did not find a suitable action principle for this theory, but we know that its partition function is a square root of the partition function we obtained for a pair of self-dual fields. As a result, its anomaly bundle $\mathscr{A}_1^\eta$ should be a square root of $\mathscr{A}^\eta$. Now $\mathscr{A}^\eta$, considered as a bundle over $\mathcal{M}/\mathcal{D}^{(2)}$ can a priori have many square roots. In fact, if the map $\mathcal{D} \rightarrow {\rm Sp}(2n,\mathbbm{Z})$ is surjective\footnote{This is always true in dimension two. We show in appendix \ref{SecCounterEx} that, at least in dimension six, there exist as well manifolds for which this happens.}, the results presented in Section \ref{SecMGPicG} shows that $(\mathscr{C}^\eta)^2$, and hence $\mathscr{A}^\eta$, admits $2^{2n(n+1)}$ square roots, corresponding to twisting the obvious square root $\mathscr{C}^\eta$ by an arbitrary character of order two.

Fortunately, the fact that we expect the anomaly bundle for a single self-dual field to be defined as well on $\mathcal{M}/\mathcal{D}^{(1,2)}$ rigidifies considerably the problem. Indeed, we saw that the character group of $\Gamma^{(1,2)}$, corresponding to flat bundles over $\mathcal{T}^{(1,2)}$, is equal to $\mathbbm{Z}_4$. This means that there is only one non-trivial character of order two. We call the corresponding flat bundle $\mathscr{T}$. It pulls back to a bundle $\tilde{\mathscr{T}}$ on $\mathcal{M}/\mathcal{D}^{(1,2)}$. We deduce therefore that either
\be
\label{EqAnomSingleSDField}
\mathscr{A}_1^\eta \simeq \tilde{\mathscr{C}}^\eta \; \quad {\rm or} \quad \mathscr{A}_1^\eta \simeq \tilde{\mathscr{C}}^\eta \otimes \tilde{\mathscr{T}} \;.
\ee
Presumably the first alternative is more natural, but strictly speaking there is no argument favoring it. 

Note that the character corresponding to $\tilde{\mathscr{T}}$ is equal to $-1$ on the elements of $\Gamma^{(1,2)}$ called anisotropic transvections and defined in the second part of Section \ref{SecMGPicG}. It was also shown there that $\mathscr{T}$ is trivial when pulled back to $\mathcal{T}^{(2)}$. Therefore there is no ambiguity as long as we consider the anomaly bundle over $\mathcal{M}/\mathcal{D}^{(2)}$.

We showed that none of the bundles $\mathscr{C}^\eta$ are isomorphic to each other, so the topological anomaly and the global anomaly of the self-dual field both depend on $\eta$. Note also that in order for the theta function to be the section of a line bundle over $\mathcal{T}^{(2)}$, $\eta$ really has to be a half-integral vector. The fact that the partition function of the self-dual field theory has to be the section of a line bundle over $\mathcal{M}/\mathcal{D}^{(2)}$ therefore imposes this condition. \footnote{Arguably, when the self-dual field considered as a building block in another theory, this condition might be relaxed, as only the partition function of the full theory should be well-defined as a section of a line bundle over $\mathcal{M}/\mathcal{D}^{(2)}$.} Its necessity was not obvious at the level of the instanton sum in Section \ref{SecIntMassless}.

\subsection{Two remarks concerning moduli spaces}

\label{SecModSpaces}

In the above, we used the fact that the restriction of the Hodge star operator on the cohomology of degree $2\ell+1$ gives us a map from the space of metrics modulo diffeomorphisms into the modular variety $\mathcal{C}/{\rm Sp}(2n,\mathbbm{Z})$ and its coverings $\mathcal{C}/\Gamma$ for $\Gamma \subset {\rm Sp}(2n,\mathbbm{Z})$. Thanks to this map, we could pull-back bundles from $\mathcal{C}/\Gamma^{(2)}$, where they are easily describable, to $\mathcal{M}/\mathcal{D}^{(2)}$. We would like now to discuss the extent to which the information gained on bundles on $\mathcal{C}/\Gamma^{(2)}$ applies to their pull-backs.

A first issue is that it could in principle be possible that some elements of ${\rm Sp}(2n,\mathbbm{Z})$ can never be realized by diffeomorphisms of a manifold $M$ of dimension $4\ell+2$. In this case there would be a strict subgroup $\Gamma$ of ${\rm Sp}(2n,\mathbbm{Z})$ containing all the group elements realized by some diffeomorphisms on some $M$. Then it could happen that there exist non-trivial bundles on $\mathcal{T}^{(2)}$ pulling back to trivial bundles over $\mathcal{C}/(\Gamma \cap \Gamma^{(2)})$. We would then see a topological anomaly where there might be none.

Fortunately this does not seem to be the case. In dimension 2, the map from the mapping class group of a Riemann surface of genus $n$ onto ${\rm Sp}(2n,\mathbbm{Z})$ is surjective \cite{Farb2010}. In higher dimension, the action of the diffeomorphism group has to preserve much more structure than just the intersection form on the middle-dimensional cohomology (for instance the intersection form on the whole cohomology...). As a result this map is in general not surjective. In dimension 6 and 14, a construction analogous to the construction of Riemann surfaces as connected sums of tori allows one to produce manifolds whose mapping class group realizes ${\rm Sp}(2n,\mathbbm{Z})$ for all $n$. This is explained in appendix \ref{SecCounterEx}. In higher dimension, the construction is still possible, but the proof fails, so strictly speaking we do not know if there are manifolds realizing ${\rm Sp}(2n,\mathbbm{Z})$ in their mapping class group.

There is a second issue. We studied bundles over certain quotients of the moduli space $\mathcal{C}$ of polarizations of the abelian variety $H^{2\ell+1}/H^{2\ell+1}_{\mathbbm{Z}}$. However, we should rather have considered the submanifold $\mathcal{M}_0$ of $\mathcal{C}$ containing the polarizations that can be realized as a Hodge star operator associated to an actual metric on $M$. In the two dimensional case, determining $\mathcal{M}_0$ is open and known as the Schottky problem \cite{Debarre1995}. From our understanding, close to nothing is known about it in higher dimension. However, even if the topology of $\mathcal{M}_0$ might be complicated, our analysis correctly describes the group of line bundles on $\mathcal{M}_0/\Gamma^{(2)}$ that pull back to trivial bundles on $\mathcal{M}_0$. Fortunately, this subgroup of the full topological Picard group contains the anomaly bundle.

%

\subsection{A holonomy formula for the pair of self-dual fields}

\label{SecHolFormPSD}

The main result of this paper was a determination of the anomaly bundle of the self-dual field theory from first principles. From the perspective of studying anomaly cancellation, however, this result is not sufficient as it stands. Indeed, we already mentioned that even if the anomaly bundle of a quantum field theory is topologically trivial, there can still exist a ``geometric'' anomaly, associated to the fact that it does not admit any natural trivialization. This anomaly is described by the curvature and holonomies of a natural connection living on the anomaly bundle. The missing piece in our knowledge about the anomalies of the self-dual field theory is the set of holonomies of this connection: the global gravitational anomaly. 

In this section, we would like to sketch how the determination of the topological anomaly achieved in this paper can be used to compute the global gravitational anomaly. We will obtain a formula for the global gravitational anomaly of a pair of self-dual fields, but not in a form very suitable to check anomaly cancellation. The case of a single self-dual field and more practical formulas will appear in another paper.

It has been known for a long time that up to a sign flip, the local anomaly for a pair of self-dual fields can be computed using the index theory of the Dirac operator coupled to chiral spinors \cite{AlvarezGaume:1983ig}. We will call the latter $D$ in the following. This fact indicates that the determinant line bundle $\mathscr{D}$ of $D$ is isomorphic to the inverse of $\mathscr{A}^\eta$ modulo torsion. We have therefore
\be
\label{EqRelAnbundle}
\mathscr{A}^\eta \simeq \mathscr{D}^{-1} \otimes \tilde{\mathscr{F}}^\eta
\ee
as bundles with connections, where $\tilde{\mathscr{F}}^\eta$ is a flat bundle over $\mathcal{M}/\mathcal{D}^{(2)}$. This is an important clue for the computation of the global anomaly. 
In \cite{Monnier:2010ww}, we showed that topologically, $\mathscr{D} \simeq \tilde{\mathscr{K}}^{-1}$, where $\tilde{\mathscr{K}}$ is the pull-back from $\mathcal{C}/\Gamma^{(2)}$ to $\mathcal{M}/\mathcal{D}^{(2)}$ of the determinant of the Hodge bundle. We also know that $\mathscr{A}^\eta$ is the pull-back to $\mathcal{M}/\mathcal{D}^{(2)}$ of the square of the theta bundle with characteristic $\eta$. As a result, the relation \eqref{EqRelAnbundle} is nothing but the pull back to $\mathcal{M}/\mathcal{D}^{(2)}$ of the relation
\be
(\mathscr{C}^\eta)^2 \simeq \mathscr{K} \otimes \mathscr{F}^\eta \;.
\ee
But in Section \ref{SecMGBun}, we computed explicitly the holonomies of $\mathscr{F}^\eta$ in terms of a character of $\Gamma^{(2)}$, see equation \eqref{EqCharCeK}. The holonomies of $\tilde{\mathscr{F}}^\eta$ are identical, so by combining \eqref{EqCharCeK} with the Bismut-Freed formula, we have a way of computing the global anomaly of a pair of self-dual fields.

The Bismut-Freed formula for the holonomy of $\mathscr{D}$ along a loop 
$\gamma$ in $\mathcal{M}/\mathcal{D}^{(2)}$ reads
\be
\label{EqHolBF}
{\rm hol}_{\mathscr{D}}(\gamma) = (-1)^{{\rm index} D} \lim_{\epsilon \rightarrow 0} \exp -\pi i (\eta_\epsilon + h_\epsilon) \;.
\ee
In this formula, we are considering the mapping torus built out of $M$ and $\gamma$, for a certain family of metrics parameterized by $\epsilon$. This family has the property that the volume of the fiber of the mapping torus shrinks to zero when $\epsilon \rightarrow 0$. From this family of metrics and $D$, one can construct a certain Dirac operator $\tilde{D}_\epsilon$, which has a kernel of dimension $h_\epsilon$. The eta invariant $\eta_\epsilon$ is defined as the value of the analytic continuation of
\be
\sum_{\lambda \in {\rm Spec} (\tilde{D})} \frac{{\rm sgn}(\lambda)}{|\lambda|^s}
\ee
at $s = 0$. From \eqref{EqHolBF}, we can deduce a holonomy formula for $\mathscr{A}^\eta$:
\be
\label{EqHolPSD}
{\rm hol}_{\mathscr{A}^\eta}(\gamma) = \chi^\eta(\gamma) (-1)^{{\rm index} D} \lim_{\epsilon \rightarrow 0} \exp \pi i (\eta_\epsilon + h_\epsilon)\;,
\ee
where $\chi^\eta$ is the character defined in \eqref{EqCharCeK}
\be
\chi^\eta(\gamma) = \exp \left( \pi i \sum_j \tilde{A}_{jj}(\gamma) - 4\pi i \sum_{i} \tilde{B}_{ii}(\gamma) (\eta_1^i)^2 - 4\pi i\sum_{i} \tilde{C}_{ii}(\gamma) (\eta_2^i)^2 \right) \;.
\ee
By a slight abuse of notation, we identified the loop $\gamma$ with the element of $\Gamma^{(2)}$ it defines through the action of the mapping class group on $H^{2\ell+1}(M,\mathbbm{Z})$.


Admittedly, it is difficult to use \eqref{EqHolPSD} as it stands in order to check anomaly cancellation, because there is no hope to compute directly the eta invariant. A more fundamental problem is that this formula describes only the global anomaly of a pair of self-dual fields. We have, in an appropriate sense, to ``take its square root'' in order to obtain the global anomaly of a single self-dual field. This is a priori an impossible task.

A first step in order to get a more practical formula is to follow \cite{Witten:1985xe} and suppose that the mapping torus bounds a $4\ell+4$-dimensional manifold. In this case, one can use the Atiyah-Patodi-Singer theorem to reexpress the eta invariant in terms of a topological invariant and the integral of a density that vanishes for theories free of local anomaly. We hope to show in a future paper that the resulting expression nicely combines with the character to yield a formula derived by Hopkins and Singer in a related context \cite{hopkins-2005-70}. The formula of Hopkins and Singer should also solve the problem of taking the square root of \eqref{EqHolPSD}.

%

\subsection*{Acknowledgments}

I would like to thank Dan Freed, Greg Moore and Boris Pioline for discussions and correspondence, as well as Daniel Persson and Boris Pioline for useful comments on a draft. Thanks also goes to the community at MathOverflow.net, especially Greg Kuperberg, Tim Perutz and Andrew Putman. Part of this work was done while I was visiting the New High Energy Theory Center at the Physics department of Rutgers University, that I would like to thank for hospitality and generous financial support. This work is supported by a Marie Curie intra-European fellowship, grant agreement number 254456.

\appendix

\section{Six-manifolds realizing $Sp(2n,\mathbbm{Z})$ in their mapping class groups}

\label{SecCounterEx}

In this appendix, we would like to show that there are manifolds $M$ of dimension $6$, for which the action of the mapping class group on $H^3(M, \mathbbm{Z})$ factorize surjectively on $Sp(2n \mathbbm{Z})$, $2n$ being the dimension of $H^3(M, \mathbbm{Z})$. \footnote{Crucial ideas for this proof were suggested by Tim Perutz on Mathoverflow.net, many thanks to him.} 

We know that this property is realized by any Riemann surface (see theorem 8.4 in \cite{Farb2010}). The idea is to construct $6$ manifolds which are as closely related as possible to Riemann surfaces. We present the construction in dimension $4\ell+2$, and specialize to $\ell = 1$ later. Recall that a Riemann surface of genus $g$ can be constructed as the connected sum of $g$ tori. Natural generalizations in dimension $4\ell + 2$ are connected sum of $n$ copies of the direct product of two $2\ell + 1$ spheres:
\be
M_{n,\ell} = \mbox{\Large $\sharp$}_{i = 1}^{n} \big(S^{2\ell+1} \times S^{2\ell + 1}\big) \;.
\ee
These manifolds are $2\ell$-connected, so by Hurewicz's theorem, we have 
\begin{align}
H^{i}(M_{n,\ell},\mathbbm{Z}) =& \; 0 \;, \quad i = 1,...,2\ell \; , \notag \\
H^{2\ell + 1}(M_{n,\ell},\mathbbm{Z}) \simeq & \; \pi_{2\ell +1}(M_{n,\ell}) = \mathbbm{Z}^{2n} \;.
\end{align}

We can construct representing cycles for the cohomology classes in $H^{2\ell + 1}(M_{n,\ell},\mathbbm{Z})$ as follows. Choose three points $x_0$, $x_1$ and $x_2$ on $S^{2\ell+1}$ together with non intersecting neighborhoods $U_0$, $U_1$ and $U_2$ of each point. When constructing the connected sums, we will arrange so that on each component $S^{2\ell+1} \times S^{2\ell + 1}$, the surgeries are performed in $U_1 \times U_1$ and $U_2 \times U_2$. In each component, $A :=\{x_0\} \times S^{2\ell+1}$ and $B := S^{2\ell+1} \times \{x_0\}$ are embedded $2\ell+1$ spheres with intersection number 1. The surgery happens outside $U_0 \times S^{2\ell+1} \cup  S^{2\ell+1} \times U_0$ so the image of $A$ and $B$ in the $i$th component give us cycles $A_i$ and $B_i$ in $M_{n,\ell}$ satisfying
$$
A_i \cap A_j = 0 \;, \quad B_i \cap B_j = 0 \;, \quad A_i \cap B_j = \delta_{ij} \;.
$$
It is also clear from this construction that the normal bundles of the cycles $A_i$ and $B_i$ are trivial.

We need now to study the diffeomorphisms of $M_{n,\ell}$ in order to compute the mapping class group. In the case of surfaces, the mapping class group is generated by Dehn twists. In fact there exist higher dimensional analogues of Dehn twists. These are compactly supported diffeomorphisms of the tangent bundle of odd-dimensional spheres, that act like the antipode on the zero section. They typically appear in the study of the monodromy of the Lefschetz degeneration (see for instance chapter 3 of \cite{Voisin2003}).

This is where we have to restrict ourself to $\ell=1$, namely the dimension 6 case. Indeed, when $\ell=1$, the cycles are 3-spheres whose tangent bundle is trivial and isomorphic to the normal bundle. In this case, we can transplant the Dehn twist in $M_{n,1}$ such that it acts by the identity outside a neighborhood of a given cycle $C$. We write $\phi^C$ for this diffeomorphism. Then its action on the homology is given by the Picard-Lefschetz formula
$$
\phi^C_\ast(D) = D + (C \cap D)C \;.
$$
Then, results obtained for surfaces show that the Picard-Lefschetz transformations generate the full integral symplectic group $Sp(2n,\mathbbm{Z})$ (see chapter 3 of \cite{Farb2010}). So the induced map from the mapping class group of $M_{n,1}$ is mapped surjectively on $Sp(2n,\mathbbm{Z})$, the result we wanted to prove.

The $d$-spheres is parallelizable only for $d = 1,3$ and $7$ \cite{Kervaire1958}, so the above construction allows one to construct manifolds with the desired property in dimension $2$, $6$ and $14$. In Section 4 of \cite{Witten:1996hc} it was shown that in the case $d = 5$, the above construction does not work: the mapping class group of $S^5 \times S^5$ factors through an index three subgroup of ${\rm Sp}(2,\mathbbm{Z})$.  
It would be interesting to know if nevertheless there exists manifolds in each dimension $2d = 4\ell+2$ whose mapping class group surjects on ${\rm Sp}(2n,\mathbbm{Z})$.

{
\small
\bibliographystyle{../../Bibliographie/BibStyle/utphys}

\providecommand{\href}[2]{#2}\begingroup\raggedright\endgroup

}

\end{document}